\documentclass[preprint,12pt]{elsarticle}

\usepackage{lineno}
\usepackage[colorlinks]{hyperref}
\usepackage{amsmath,amssymb,amsfonts}
\usepackage{graphicx}
\usepackage{caption}
\usepackage{subcaption}
\usepackage{booktabs}
\usepackage{comment}
\usepackage{geometry}
\usepackage{cancel}
\usepackage[percent]{overpic}
\usepackage{parskip}
\usepackage{footmisc}
\usepackage{fancyhdr}
\usepackage{algorithmic}
\usepackage{textcomp}
\usepackage{xcolor}

\usepackage{mdframed}   

\usepackage{tikz}
\usetikzlibrary{shadows.blur,positioning,shapes.geometric, decorations.pathreplacing}

\newmdenv[
  linecolor=white,
  linewidth=0.5pt,
  backgroundcolor=white,
  skipabove=20pt,
  skipbelow=3pt
]{footer}

\begin{document}

\begin{frontmatter}

\title{Smart On-Street Parking: Survey of Actual Implementations in Cities and Insights from Practitioners
}

\author[inst1]{Enock NDUNDA\corref{cor1}}
\ead{enock.ndunda@univ-lyon1.fr}

\author[inst1]{Alexandre NICOLAS\corref{cor1}}
\ead{alexandre.nicolas@cnrs.fr}

\cortext[cor1]{Corresponding authors}

\affiliation[inst1]{organization={Universite Claude Bernard Lyon 1, CNRS France},
            addressline={Institut Lumière
Matière}, 
            city={Lyon},
            postcode={69100}, 
            state={Villeurbanne},
            country={France}}

\begin{abstract}
Smart solutions for on-street parking, which collect and leverage real-time information about on-street parking space availability to guide drivers or adjust policies, have attracted considerable attention in academia and in the corporate world,
but comprehensive feedback on actual implementations was still missing. Here, we survey around 25 smart parking (SP) implementations in cities across the world using online sources. To get more candid insights, we complement this objective review with case studies centred around  interviews  that we conducted with practitioners from ten cities across continents (San Francisco, Saint Pete Beach, Penang, Douala, Soissons, Grand Paris Seine Ouest, Montpellier, Frauenfeld, Zurich, Perth).
Summing up our observations, we underline the broad diversity of SP implementations in terms of contexts and scales, from 2-to-3-year small-scale pilot studies to large deployments that take centre stage in a city's mobility policy. Technological choices also vary widely, from the ground sensors used in pioneering deployments in the early 2010s and still in use, to static cameras (which cover more spaces per device), and to mobile cameras with automatic licence-plate recognition embarked in roaming cars, a more and more popular solution for parking control. Different attitudes to the role given to smartphone applications are also noticed. But, importantly, not only means, but also goals differ: facilitating parking control and enhancing revenue, or providing data for a curb-pricing strategy, or feeding live data into navigation algorithms to reduce parking search times. Unfortunately, their level of achievement is seldom gauged with robust metrics. Hardware durability issues are mentioned as causes of premature termination, particularly for `first-generation' ground sensors, but so, too, are fluctuating political will and changing priorities. Smaller-scale, geographically isolated implementations and pilots are particularly vulnerable to these fluctuations, to discontinued funding or defaulting start-ups, and to limited public awareness.
\end{abstract}

\begin{keyword}
On-street parking; smart parking systems; smart city; implementations; IoT
\end{keyword}

\end{frontmatter}

\section{Introduction}\label{intro}

As urbanization continues to rise globally, more efficient parking management becomes critical in more and more cities.
Growing urban population, often rising car ownership levels, and strong car dependence conspire with (necessarily) limited parking supply to intensify congestion in city centres, prolong parking search times, and contribute to aggravating pollutant emissions \cite{shoup2018parking}. Many car trips end on a parking search phase (which lasts a couple of minutes on average in e.g. Frankfurt, Germany, but with large fluctuations \cite{saki2024cruising}), and the associated cruising traffic represents a significant share of the total traffic (evaluated to e.g. 15\% in Stuttgart, Germany \cite{hampshire2018share}).
To alleviate these externalities, previous studies have highlighted the need for sustainable and adaptive solutions, when the public transportation network is not competitive \cite{luke2018car} and active mobility is not sufficiently widespread. These solutions include parking guidance systems directing drivers to non-saturated parking facilities \cite{axhausen1993effectiveness}, demand-based pricing that adjusts curb parking rates to reduce the occupancy on the busiest block faces, possibly in a dynamic way \cite{pierce2013getting}, digitally accessible real-time information communicated to drivers \cite{lin2017survey}, as well as the possibility to reserve one's spot in advance in some parking lots (e.g., in the Bay Area, near San Francisco, United States), notably for park-and-ride users \cite{shaheen2005smart}.

Many of these `smart' solutions have become commonplace. Systems that have been around for decades include parking guidance to, and in, \emph{off-street} facilities with the help of variable message signs (VMS) \cite{axhausen1993effectiveness} and/or red or green lights above parking spaces depending on their availability. Online marketplaces where one can book a (typically private and off-street) parking space for a limited duration (\emph{gaparking} \cite{lin2017survey}) have also flourished, with a considerable number of start-ups and companies that purport to connect space owners and drivers. Finally, as smartphones have become ubiquitous, so have smart (mobile-based) payment methods.

Comparatively, smart \emph{on-street} parking solutions (SPS) relying on real-time space monitoring have not expanded as far and wide globally. Admittedly, many start-ups have blossomed to help drivers find a spot on the fly in busy downtown areas notably via a smartphone application and/or help municipalities monitor on-street parking occupancy and control parking payment. In parallel, the academic literature has also seen proposals galore for sensor technologies that can operate outdoors \cite{paidi2018smart}, for wireless methods to pass parking-related information, and for algorithms that predict parking occupancy based on non-exhaustive data \cite{evenepoel2014street,bock2019smart}, efficiently allocate parking spaces, and guide drivers towards them \cite{aydin2017navigation}, as well as for economical city-scale system deployment strategies \cite{lin2015urban}. Such is the number of publications on the topic that several review papers have tried to summarise them, either with an all-encompassing scope fashion \cite{lin2017survey} or a more specific focus on one or several facets: all types of actual smart parking implementations before 2005 \cite{shaheen2005smart}, systems design and AI  \cite{cahyadi2023literature},  sensors and related technology \cite{idris2009car,barriga2019smart} (where one can read that already in 2009 "vehicle detections sensors [are] aplenty on the market"), and a close-up on the challenges associated with outdoor use \cite{paidi2018smart}. 

In this light, is it worth piling yet another manuscript atop the existing literature? We contend that the translation of the foregoing ideas and services into smart-parking (SP) deployments \emph{in the real world} actually deserves careful examination. Indeed, in the complex social, technical, and political environment of cities, this translational stage involves multiple hurdles, from the accuracy of sensors in real urban settings to their long-term maintenance and to the economic viability of their vendors (often small companies) and to the question of the persistence of the political support for the initiative. There may also be a gap between expected efficiencies and on-the-ground reality. As we will see, even SP projects widely highlighted as `success stories' come along with some caveats and nuances. In addition, the generic goals pursued by SP, namely
\begin{enumerate}
    \item to shorten drivers' parking search times,
    \item to improve user experience through accurate, reliable, and adaptive information services,
    \item to reduce congestion in city centres by limiting cruising traffic, and
    \item to enhance parking revenues in a cost-effective way, 
\end{enumerate}
are weighted quite differently depending on the context; it thus makes little sense to gauge success with just one metric.

Accordingly, this paper targets actual implementations of on-street SPS in cities around the world. Not only does it review the technologies in use, project sizes, and effectiveness in the different cities, but it also aims to provide critical insights into the diverse ambitious and practical challenges, and the balance between technological promises and actual limitations. For this purpose, we have reviewed around 25 actual SPS implementations that give (quasi-)real-time information on on-street parking availability. We complement this objective and fairly comprehensive review with semi-structured interviews conducted with practitioners from ten cities across continents to get more candid feedback and personal perspectives on the different facets of SPS.

The remainder of the paper is organised as follows. Chapter \ref{methods} introduces and delineates the types of SPS under consideration, in terms of technology and deployment, and exposes our methodological approach, including data sources, analytical tools, and limitations. Chapter \ref{results} reports the results of our systematic survey, which highlights the evolution and growth of SP initiatives globally and reviews actual implementations of on-street SPS around the globe. Chapter \ref{casestudiesChap} focuses on specific implementations selected as case studies, with more candid, all-around insights notably gained from interviews with stakeholders and practitioners. Finally, Chapter \ref{discussion} summarises our findings and discusses the effectiveness of SPSs across key performance indicators such as congestion and search times, parking pricing and revenue, and user satisfaction, and it underlines common vulnerabilities in SP deployments.

\section{Definitions \& Methods}\label{methods}

\begin{figure}[htbp]
\centering
\includegraphics[width=0.5\textwidth]{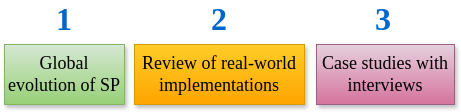}

\caption{Steps followed in this paper.}
\label{methodoverview}
\end{figure}

To clarify the scope of our study, we start by classifying SPS based on their functionality, effectiveness and target users.  Then we describe our methodological choices regarding  the literature review of academic and industrial publications, the systematic analysis of real-world  SP implementations, and the set of semi-structured expert interviews with stakeholders. Figure \ref{methodoverview} gives an overview of these steps.

\subsection{Delineation and Classification of Smart Parking Solutions}

SP encompasses a wide range of solutions to facilitate parking, from the basic payment process to complex, sensor driven, AI-augmented ecosystems. We delineate three broad categories: Payment Systems, Detection \& Monitoring Systems, and Integrated Smart Platforms.

\subsubsection{Payment Systems}\label{paymentsystem}

\begin{itemize}
    \item [A.] \textbf{Traditional Parking Meters} are stand-alone, coin-operated or card-enabled meters. This long-standing form of parking control \cite{jacobson1950limitations} is still widespread in many urban environments thanks to its low cost and simplicity.

    \item [B.] \textbf{Smart Parking Meters} (see Figure \ref{fig1}) extend conventional meters with wireless connectivity, digital payment capabilities (i.e., credit cards, e-wallets and SMS), and limited data reporting (transceiver functionality). They often interface with mobile applications or central servers, enabling zone-based enforcement and basic occupancy estimation through transaction logs \cite{YANG2017165,jprdondaniel}. Though not directly measuring space usage, their digital footprint provides valuable temporal usage patterns.

    \item [C.] \textbf{Mobile Parking Payment Applications} are software-only solutions enabling users to pay for parking on their smartphone. They can enhance user convenience and reduce meter infrastructure, but they do not provide direct occupancy data, unless coupled with auxiliary sensing architecture \cite{ghezzi2010mobile}. 
\end{itemize}

Since our focus is on \emph{real-time information}, these payment systems will not qualify as SPS in the following.

\subsubsection{Occupancy Detection Systems}\label{detectionmornitoring}
\begin{itemize}
    \item [A.] \textbf{Overhead and Ground Sensors} are fixed sensors which are buried in the ground underneath parking spaces, fixed on the ground, or mounted overhead, above the parking space (depending on their type). They rely on various physical waves to detect the presence of a car: ultrasound (ultrasonic sensors), radio waves (radar), infrared, visible light (optical sensors), local magnetic field variations (magnetometers), etc. These diverse technologies have been largely documented in past publications \cite{shaheen2005smart,idris2009car,lin2017survey,paidi2018smart,barriga2019smart}. Magnetometers (i.e., inductive coils that detect the variations of the local magnetic flux when cars - more precisely, their metal parts - are above them)  and ultrasonic sensors (which sound out ultrasounds and capture the reflected signals) compete for the title of most popular solution: magnetometers are reported to be the most common devices for municipal deployment (as of 2017) in \cite{lin2017survey}, whereas ultrasonic sensors seem to dominate the academic publications reviewed in \cite{barriga2019smart} and may be even more cost-effective \cite{lin2017survey}. (Costs should not be estimated only based on the cost of the device, though, as installation, maintenance, and durability also matter for cost effectiveness and in several municipal SP projects equipment-related costs were bundled in a broader package or subsumed into monthly subscription fees; the financial question will be one facet of the case studies of Chap.~\ref{casestudiesChap}). 
    On the whole, ground sensors offer high per-space resolution but require extensive installation and maintenance; reported detection accuracies range from 85–98\% \cite{sfpark_2024}, varying with environmental conditions and time (see Chap.~\ref{casestudiesChap}). 
    It is noteworthy that outdoor conditions are much more challenging than indoor parking \cite{paidi2018smart}, because of sunlight, rainfall, the absence of a ceiling, among other adverse factors.

    \item [B.] \textbf{Static or mobile cameras}, coupled with AI-powered object recognition software, are increasingly used to monitor occupancy. These vision-based systems rely on either static surveillance cameras  or mobile platforms such as automatic number-plate recognition (ANPR) vehicles. Static cameras use computer vision to detect vehicle presence in delineated parking spaces and estimate near real-time occupancy, often integrated with AI for license plate verification. Mobile ANPR systems enable high-coverage enforcement through periodic sweeps; this technology was initially deployed to address illegal parking, but it has diversified. Their accuracy hinges on the capturing angle, the camera quality, lighting, how well the algorithm was calibrated, and possibly also  license plate neatness; it often exceeds 90\% in ideal conditions. Finally, these systems may raise some privacy concerns due to the video-recording (and the frequent transmission of images to confirm that a car is illegally parked).
\end{itemize}

\begin{figure}[h]
    \footnotesize
    \centering
    \centering
    \includegraphics[width=0.45\linewidth]{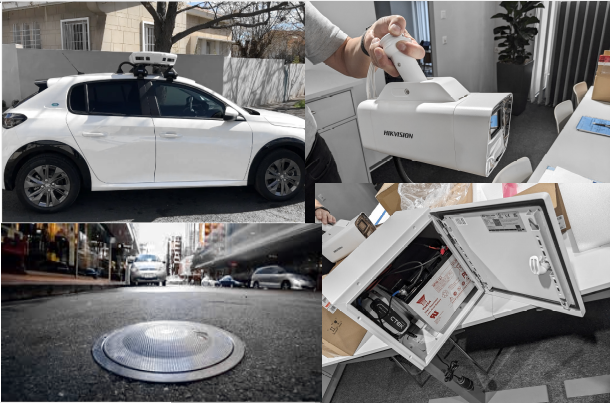}
    \caption{A variety of SPS: Pictures of smart parking cameras \& sensors.}
    \label{fig1}
\end{figure}

Figure \ref{fig1} shows some examples of ground sensors, mobile cameras, smart parking meters, and static cameras.

\subsubsection{Integrated Smart Platforms (ISP)}\label{integrated}

\begin{itemize}
    \item [A.] \textbf{Real-time information on on-street parking availability}, based on the occupancy detection systems, can be communicated to the general driver typically via a smartphone application or accessible only to the City or transportation authority for monitoring (via a dashboard) and payment control purposes. Apart from cases for which full accuracy is critical (e.g., to control payment and issue fines), non-exhaustive real-time data can be completed by occupancy predictions through  machine-learning, which leverage  historical data and behavioural analytics. The improvement of these predictions has been the target of several academic works \cite{yang2019deep,zhao2020comparative,vakrinou2025leveraging}. For instance, Zhao et al. found that support-vector machines  perform well for occupancy prediction for car parks of variable sizes \cite{zhao2020comparative}, while Yang and colleagues reached a mean absolute percentage error (MAPE) of 10.6\% for block-level occupancy forecasting 30 minutes in advance \cite{yang2019deep} and Vlahogianni's team arrived at a MAPE below 9\% for 1-hour-in-advance predictions of off-street parking occupancy with an LSTM featuring attention models, by also including traffic data as input. Beyond academia, companies have also made substantial investments to develop AI-powered occupancy prediction capabilities. For example, ParkNav claims to provide their application users with predictions that are at least ``80 to 90\%'' accurate.

    \item[B.] \textbf{Smart navigation} assistance towards available parking spaces can also be provided to drivers, based on real-time occupancy information. The principle is similar to a traditional GPS navigation system, except that instead of guiding the driver to their final destination the system guides them to a nearby location where they are likely to find vacant parking spots. This service is often included in smartphone applications that display real-time parking availability. In some sense, this service can be regarded as an upgraded and individualised version of the variable message signs that inform drivers on availability in nearby parking facilities and that have been in place for more than thirty years (e.g., in London, Nottingham and Frankfurt-am-Main \cite{axhausen1993effectiveness}).
    
    \item[C.] \textbf{Digital parking reservation} consists in the possibility to book a parking space online in advance. This option is widespread for privately owned parking spaces that are rented in online `marketplaces' developed by companies (\emph{gaparking}) and it was implemented early in some off-street parking facilities where finding a space is critical (e.g., at park-and-ride facilities, such as the Bay Area Rapid Transit in California, where spots could be reserved for a single day or on a longer term, as recalled by Shaheen et al. \cite{shaheen2005smart}). On the other hand, for municipally managed on-street parking, none of our detailed case studies (see Chap.~\ref{casestudiesChap}) includes this reservation option.
    
    \item [D.] \textbf{Comprehensive  SP Applications (Payment, Navigation, Reservation)} integrate a combination of SP services. For instance, these mobile platforms can be used as payment systems as well as for real-time navigation. In that case, they leverage real-time occupancy data, often fused from multiple sources (sensors, historical data, traffic feeds). Users are guided to available spots, thus spending less time searching for parking and allegedly alleviating congestion as a result. These solutions embody the convergence of IoT, cloud computing, and user-centric design as shown in Table \ref{table1}. Our case studies (Chap.~\ref{casestudiesChap}) have revealed that, because of the rapidly evolving ecosystem of start-ups in the parking business, the vendors (service providers) may change in the course of a project and diverse mobile applications may be put forward (either successively or, in some cases, simultaneously), which makes it more difficult to get drivers to be aware and use this mobile service. Having different parking-related services (and beyond) combined in a single application can help mitigate this problem.
\end{itemize}

Table \ref{table1}, inspired from \cite{lin2017survey}, gives an overview of the hardware and software associated with the above smart parking  solutions, along with their degree of innovcation and cost (i.e installation, maintenance, \& operational costs).

\begin{table}[h]
    \footnotesize
    \centering
    \begin{tabular}{|p{4.4cm}|p{4cm}|p{3.4cm}|p{1.5cm}|p{1.5cm}|} \hline
        \textbf{Solution} & \textbf{Hardware} & \textbf{Software} & \textbf{How innovative?} & \textbf{How costly?} \\ \hline
        Traditional parking meters & Mechanical coin/card meter & -& + & Variable \\ \hline

        Smart parking meters & Connected digital meter & Payment gateway & ++ & Variable \\ \hline

        Smart Payment Apps & Smartphone, cloud-based server & App, cloud backend, API & + & + \\ \hline

        Ground sensors & In-ground magnetometers, ultrasonic sensors & Sensor network, edge processing, API & +++ & +++ \\ \hline

        Overhead sensors & Infrared, radar, LiDAR sensors & Image processing, signal filters & +++ & +++ \\ \hline

        Static and mobile cameras & Fixed/vehicle-mounted cameras + ANPR & CV + OCR + real-time analytics &+++ & +++\\ \hline

        Smart Parking Apps & Smartphones, server, 3rd-party integration & App, machine-learning prediction, navigation & +++ & ++ \\ \hline
    \end{tabular}
    \caption{Delineation of Smart Parking Solutions Based on Technology, Efficiency, and Cost. Costs give an idea of the full deployment cost of the solution.}
    \label{table1}
\end{table}

\subsection{Methods used to gauge academic trends}\label{method1}

To assess the academic trends relative to SP, this paper relies on four data sources: \emph{Google Scholar}, \emph{IEEE Xplore}, \emph{ScienceDirect} and \emph{Springer Link}. We used the following keywords and phrases in the search engines to find relevant papers:  "smart parking services", "parking meters", "smart parking systems", "autonomous parking", "smart parking solution", "smart car parking", "intelligent parking", "IoT parking", "automated parking",
    "connected parking", "parking management system", "parking wireless sensor",    "parking" system for vehicles", "smart parking reservation", "parking using image processing". 
The results were collected and screened for candidate publications.

\subsection{Methods used to review SP implementations}\label{method2}
To capture a global perspective on \emph{implemented} SPS, we conducted a systematic review of cities that have implemented such technologies. We did our best to reflect geographic diversity (even though language has probably generated a selection bias) and to include multiple types of SP technologies.

Our first set of analysed implementations included 51 cities from 21 countries and 6 continents. Further screening reduced the total to 25 cities, after excluding implementations that only offer smart payment methods or smart parking meters. Importantly, not all these solutions are still active: a number of them have been terminated; their study is interesting, to avoid a focus on only viable or successful projects. Figure \ref{map} shows the geographic distribution of surveyed cities on the world map, distinguishing between implementations passed in review and detailed case studies.


\subsection{Interviews of stakeholders: methodology}\label{method3}
For more direct feedback, following in the steps of \cite{brooke2017street}, we complemented the foregoing global review with a series of semi-structured interviews with stakeholders. The objective was to validate and enrich the findings of the systematic review by capturing real-world experiences, contextual nuances, and expert insights.

\begin{figure}[h]
    \centering
   \includegraphics[width=0.7\textwidth]{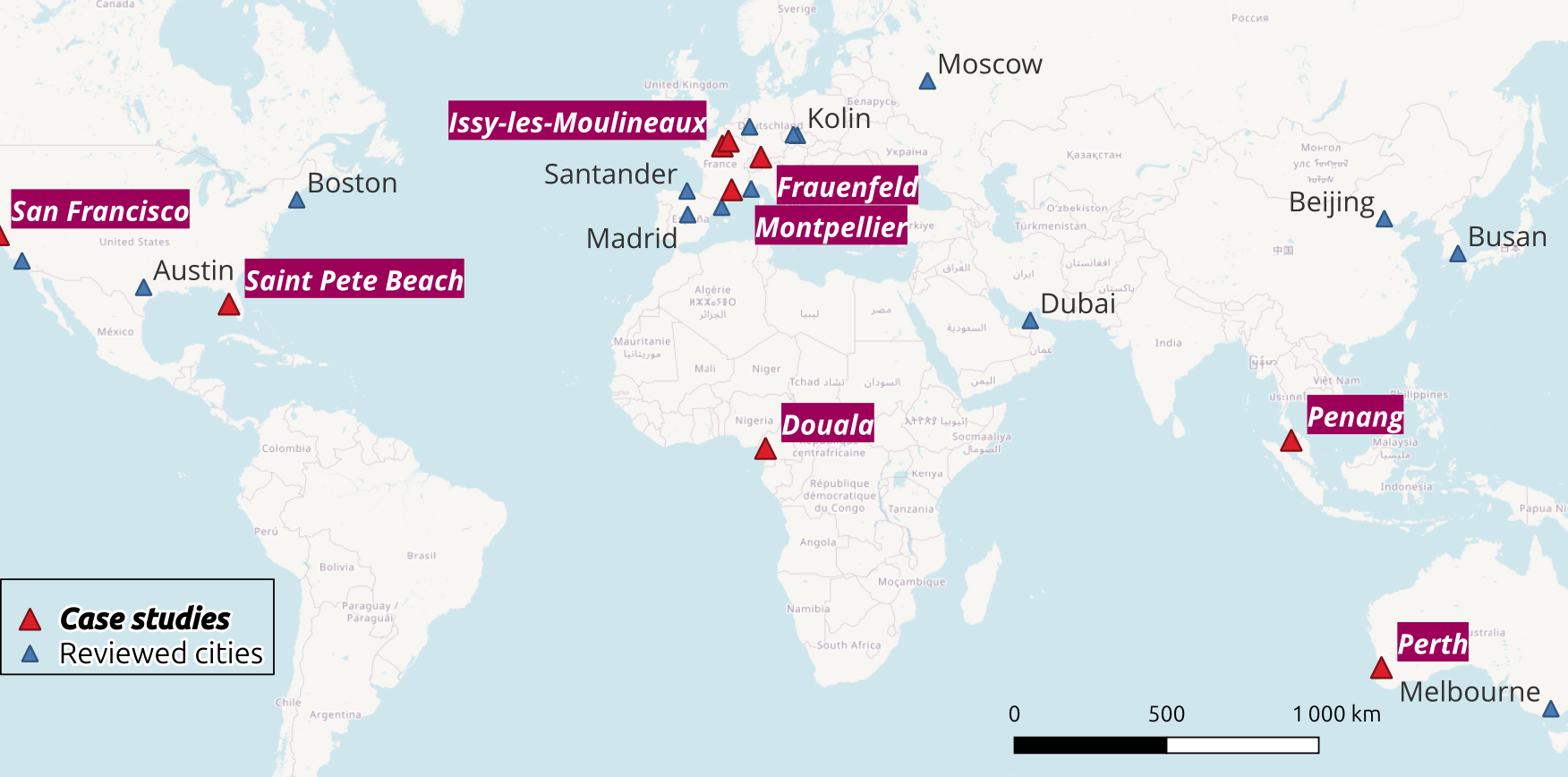}
 \caption{Geographic distribution of the cities whose SP implementations were globally reviewed and/or used as case studies.}
    \label{map}
\end{figure}

For that purpose, we contacted stakeholders from various cities around the world where smart on-street parking efforts had been carried out. Our focus was on executives in transport departments or in municipal technical services, but we also approached some SP companies' Chief Executive Officers. Our contact endeavours (mostly by email) yielded a response rate of about 40\%.
The selected interviewees represent a diverse geographic (see Fig.~\ref{map}) and technological spectrum.

Interviews were conducted in the form of 50-minute to 1-hour long video calls with two interviewers (the authors) and generally one interviewee. Before each interview, we got acquainted with the city of interest and prepared a short list of open questions, most of which were similar across cities:
\begin{itemize}
    \item ``What motivated the introduction of the smart-parking solution?'' / ``Can you introduce us to the context of the SP implementation?''
    \item ``How reliable was the technological solution?''
    \item ``Were traffic and parking search conditions quantitatively evaluated before SP implementation?''
    \item ``Were there objections to the implementation?''
    \item ``How was the implementation appraised by citizens?''
\end{itemize}
Other questions were specific. But we mostly insisted that the interviewee should have time to dwell on the facets that they considered most relevant. We also made clear that they could speak openly and, should they have accidentally divulged a piece of information that they would prefer to retain, they could write to us later and ask us to not mention it. After finalising the first version of our manuscript, we sent it to the Interviewees and left them two weeks to go through it and, if need be, ask for amendments; a couple of them responded, requesting a few corrections of factual errors on numbers or assertions.

\newpage


\section{Global review of smart-parking implementations: Results}
\label{results} 

This chapter exposes the results of our SP-related investigations, first by considering SP trends over time, then analysing the characteristics of SP implementations over the world.

\subsection{Growing academic and industrial interests for SPS}

\begin{figure}[h]
    \footnotesize
    \centering
\begin{subfigure}[t]{0.499\textwidth}
    \footnotesize
    \centering
    \includegraphics[width=0.9\textwidth]{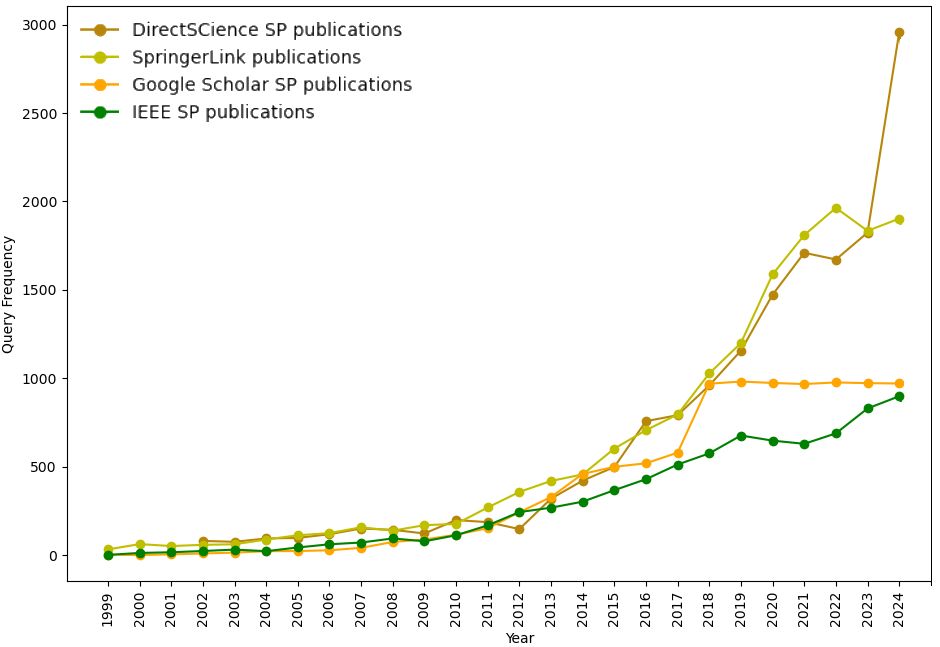}
    \caption{Annual SP Publications}
    \label{publications}
\end{subfigure}
\hfill
\begin{subfigure}[t]{0.49\textwidth}
    \footnotesize
    \centering
    \includegraphics[width=0.9\textwidth]{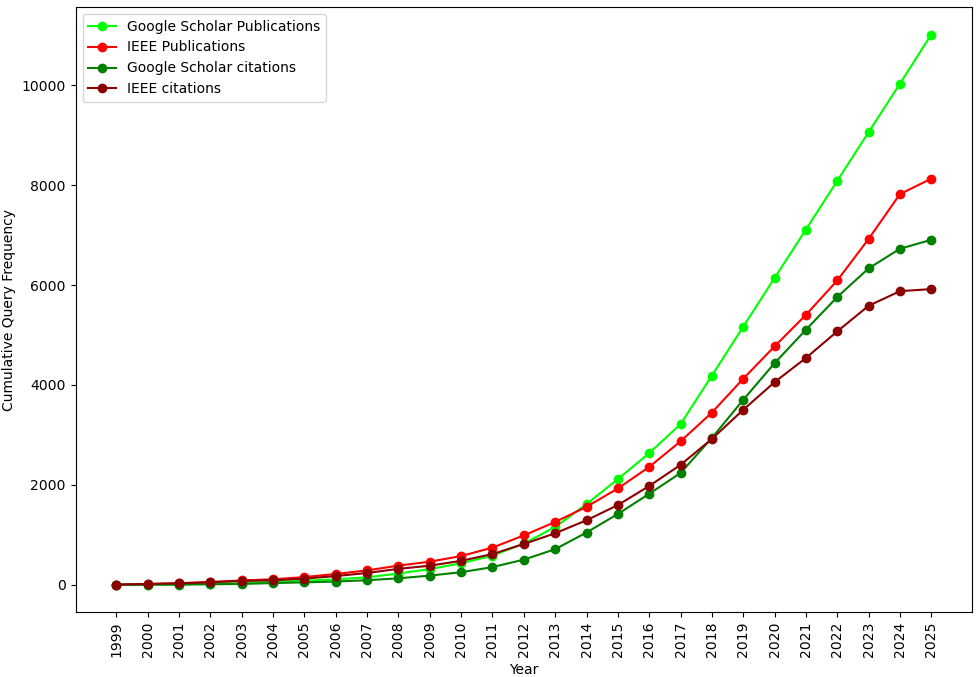}
    \caption{Cumulative Annual Citations}
    \label{cum_cite}
\end{subfigure}
    \caption{Evolution of the academic literature about SPS between 1999 and 2025 (July, 2nd)}
    \label{sp_metrics}
\end{figure}

We start by assessing the evolution of the academic interest in SP, following the search method described in Sec.~\ref{method1}. Figure~\ref{sp_metrics} evinces that this interest has been steadily rising since year 2000, as reflected by both the number of publications per year on the subject and the number of citations recorded in the different databases under consideration.
Until 2005$\sim$2010, the literature on the topic remained modest, but strongly expanded after 2010. This rise aligns with the global push towards smart city initiatives, especially in urban mobility and intelligent transportation systems, echoed by heavy investment in smart infrastructure, as well as IoT-enabled parking solutions, such as SFPark funding by the Federal Highway Administration in 2010, Texas 2015 Regional Truck Parking Information and Management System across eight states (\$25 million), the \$6.9 million grant by  the U.S. Department of Transportation for the I-10 Corridor Coalition Truck Parking Availability System (for trucks) started in 2020 in the USA, and QPARK, a Horizon 2020-funded E.U. project led by Ubiwhere to develop low-cost smart parking sensors and analytics, in the E.U.
Concurrently, along with the swift pervasion of smartphones since the early 2010s, technological advancements such as low-power sensors \cite{sensors2008}, GPS, cloud computing \cite{armbrust2010view}, and the rise of mobile apps made SPS feasible and scalable. 

 In Fig.~\ref{sp_metrics}, an inflection in academic publications starting around 2020 might be noticeable. This tentative slowing down might reflect a shift in research priorities to broader or newer urban mobility challenges such as electric vehicle (EV) integration, multimodal transport, or more sustainable alternatives.  These trends reveal how closely SP research has tracked urban policy, technology readiness, and funding cycles.

\newpage

\begin{figure}[h!]
    \footnotesize
    \centering
        \includegraphics[height=0.9\textheight]{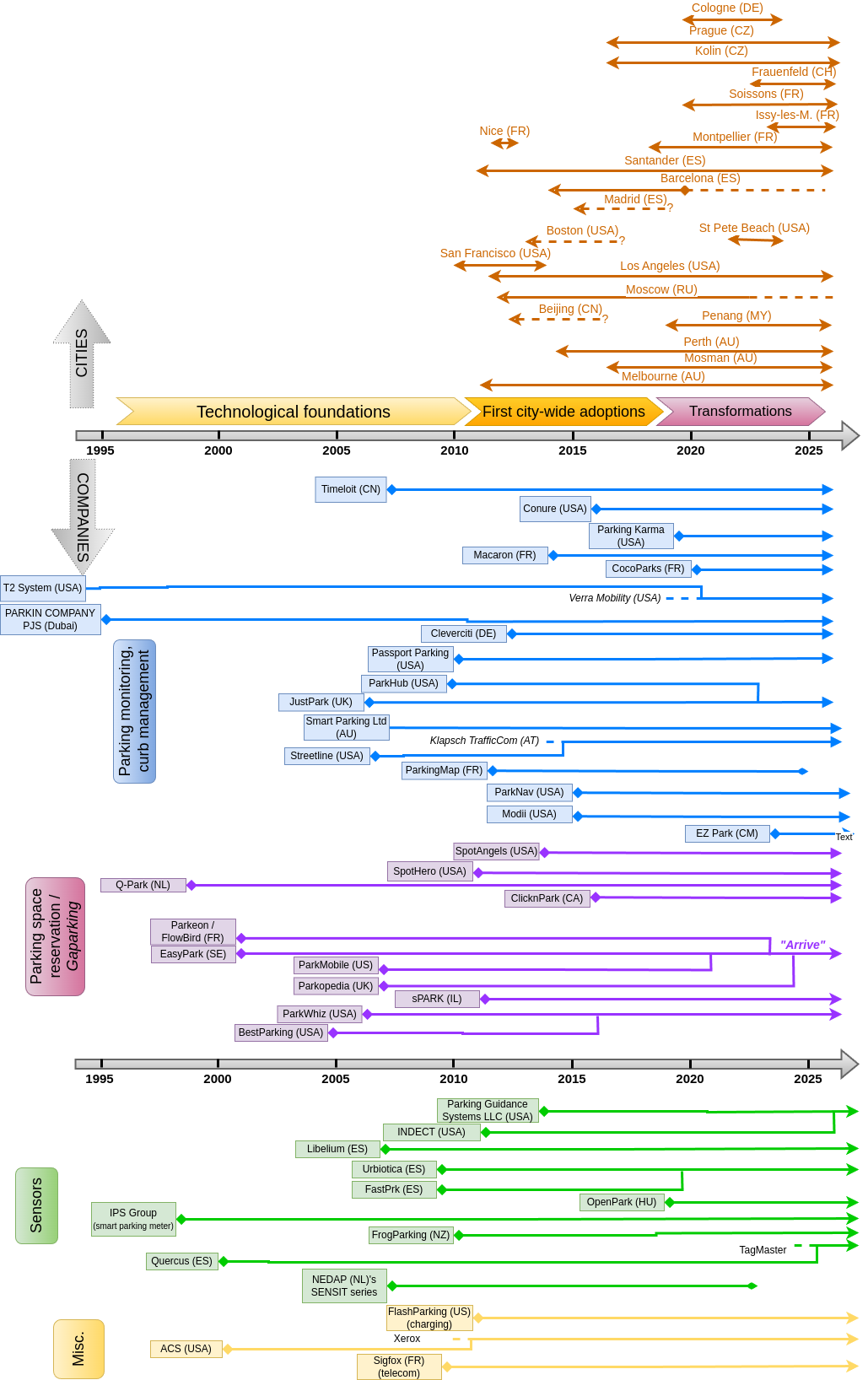}
    \caption{Evolution of Service Providers and Milestones. The thick coloured lines start at the foundation of the company or the first implementation and terminate at its end. There may occasionally be an uncertainty of one or two years about these dates.}
    \label{startupsmilestones}
\end{figure}

\begin{figure}[h]
    \footnotesize
    \centering
    \includegraphics[width=0.99\linewidth]{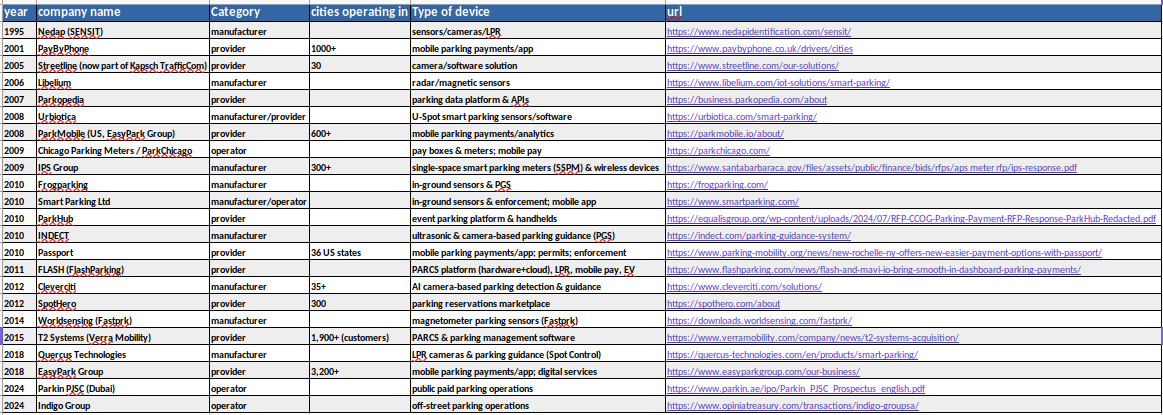}
    \caption{Details of the activities of some companies involved in SP, with website references.}
    \label{startupsgrowth}
\end{figure}

The extended timeline displayed in Fig.~\ref{startupsmilestones} includes both city-scale on-street SP implementations, in its upper part, and major companies involved in SP (also refer to Fig.~\ref{startupsgrowth} for further information and references about start-ups and companies involved in the SP market). While some of these entities burgeoned within big firms, many of them (a large majority of those shown in the figure) emerged as start-ups.
Since SP implementation is a multi-faceted activity, it is worth classifying these companies by their main activity; in  Fig.~\ref{startupsmilestones}, we considered the following categories: general parking monitoring or curb management, digital parking reservation and \emph{gaparking} (often with a strong off-street activity), sensor development and production, other specialisation.  
Admittedly, a substantial fraction of the companies span boundaries between activity sectors, but many have a core speciality, which explains why cities often commission several of these firms to design and deploy their SP strategy, as we will see  in Chap.~\ref{casestudiesChap}.

Even though one should acknowledge a probable bias towards firms and SP implementations that have survived until the present date, the timeline clearly highlights the existence of different phases of development. Many sensor (hardware)-making companies were founded in the 2000s, notably in Europe, a period to which we refer as the times of the `technological foundations', while the SP service providers that still dominate the market typically emerged between 2005 and 2015. Meanwhile, city-wide on-street SP implementations started to blossom in the 2010s (with the much-publicised SFPark project in the early 2010s, in particular). After this early phase of large-scale adoptions, a rising number of cities (in Europe, in the USA, in Australia, among other places) came to test on-street SP, often first in the vicinity of University campuses or in well-off districts, without necessarily adopting it then on a larger scale or on a permanent basis. 
In a business sector where there is a premium to well established actors (because they can build on experience and notoriety, and streamline costs and risks), a consolidation stage was to be expected and, indeed, the SP market started to consolidate around 2020, with several mergers between significant market players (see the merging branches in Fig.~\ref{startupsmilestones}). In parallel, and possibly in a related way, along with the evolution of the  political focuses (e.g., with a stronger will to move away from the car-centric city paradigm in Europe \cite{kodransky2010europe}), technological transformations were taking place, spurred on by the boom of AI and computer vision.

\subsection{Regional distribution of SP implementations}

Let us now consider how SP implementations are distributed around the world. A caveat is however in order, before any comment: the ``baseline'' on-street parking situation differs widely between cities and continents. While on-street parking is dominant in cities such as Los Angeles (75\% of all parking spaces) and San Francisco (63\%) on the West Coast of the USA, it is less abundant than off-street parking in Europe (37\% on average), and generally scarce in Eastern Asian cities such as Beijing (5\%) or Tokyo, as highlighted by Lin et al. \cite{lin2017survey}. This is naturally reflected in the contrasted figures for parking coverage across cities, which represents  up to 81\% of land use in Los Angeles as of 2015 and as little as 7\% in Tokyo. One should bear in mind this diversity when commenting on municipal SPS deployments.

The geographic distribution of cities included in our extensive SPS review  or in our specific case studies is shown in 
Fig.~\ref{map}. Europe and North America dominate our collation of SP implementations. Although we aimed for as uniform a coverage as possible, this dominance possibly stems from some bias in the search focus, language, and the diverse mediatisation of such initiatives; it probably also reflects the larger number of adoptions or tests of SP strategies that actually have taken place. 

In the Southern hemisphere, Australia has been an active SP territory. In contrast, the African continent (where several cities that develop quite fast and in a partly uncontrolled way are increasingly confronted with parking difficulties) only contributes one or two cases, with a solution launched in 2024; nevertheless, in Sec.~\ref{casestudiesChap}  a case study on an emerging solution in a couple of African capitals will be exposed.

\subsection{Coverage of SP implementations}
\begin{figure}[h]
    \footnotesize
    \centering
    \includegraphics[width=0.70\linewidth]{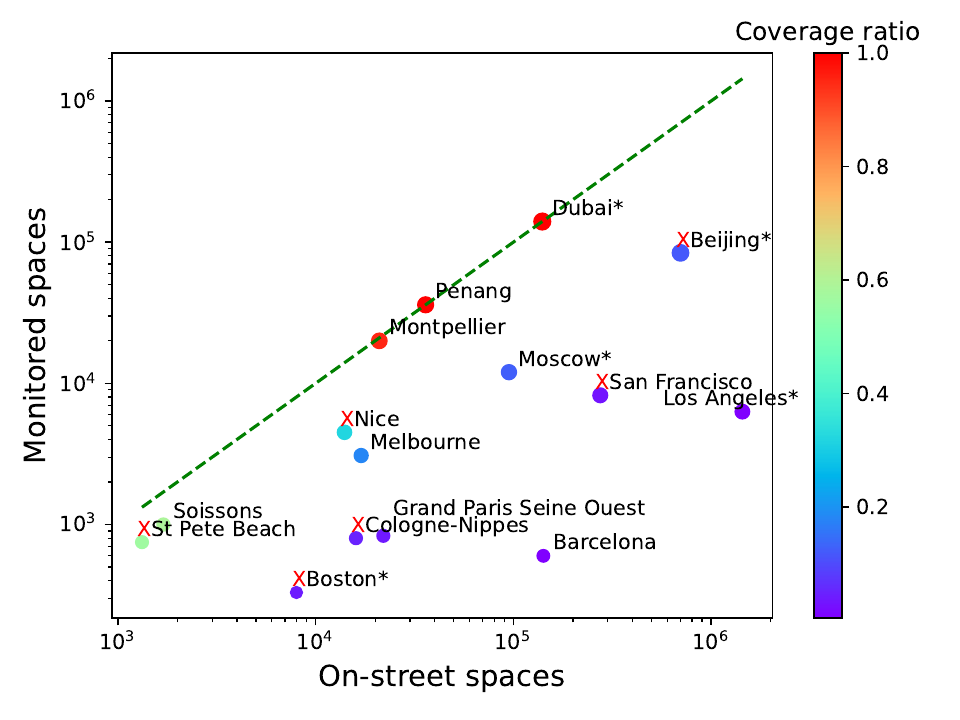}
    \caption{Number of on-street parking spaces covered by the SP system as a function of the total number of on-street spaces, for various cities included in the review. The dashed green line, representing the line $y=x$, corresponds to full coverage. Circle sizes are proportional to the logarithm of the number of covered spots. A red star before the city's name indicates a terminated project.\\
    $^{\star}$ For the cities marked with a star, the total number of spaces is uncertain, despite email requests to the transport departments. For Boston, we use the estimate (8,000) indicated to us by the City of Boston, even though it is highly dubious. For Los Angeles (LA), we made the assumption that the numbers of spaces in LA City and LA County are proportional to their respective populations. For Moscow, the monitored spaces do not include those observed by ANPR cars.}
    \label{implemented_solutions_year}
\end{figure}

The examples shown in Fig.~\ref{map} include cities of very different sizes and SP deployments that cover wide-ranging fractions of the total number of on-street spots.
 This diversity is reflected in Fig.~\ref{implemented_solutions_year}, which plots the number of spaces monitored by an SPS against the (estimated) total number of on-street spaces.

Not surprisingly, irrespective of the size of their on-street parking supply, the cities included in the SP review are generally known historically for their congested traffic (Boston, Moscow, San Francisco, Barcelona, etc.) and for a perceived shortage of parking supply compared to the parking demand. Parking-related policies are often deemed pivotal in these cities to alleviate the traffic conditions.
For instance, the \emph{SFPark} project targeted some of the busiest parking areas  in San Francisco and, by dynamically adjusting parking rates in each street segment, it aimed to homogenise the occupancy and to reduce parking search by curbing the occupancy on the busiest street portions \cite{sfp_evaluation}. The case of Moscow is also interesting in this respect, as on-street parking used to be free until 2012, but the city then gradually turned to paid parking with rapidly increasing parking rates, in order to reduce car traffic and promote car-sharing. 

One could thus think that the SPS coverage ratio, i.e., the fraction of spaces covered by the SPS (highlighted by the colour of the symbols in  Fig.~\ref{map}) depends on how congested traffic is known to be in the city. This would be wrong. In fact, since one of the goals of municipal SP implementations is to generate parking revenue by controlling payment, SP is rarely deployed for free spaces, where no additional revenue can even up SP operation, and the percentage of free spaces varies widely across cities (and along time, as we saw in the example of Moscow). Besides this effect, the fraction of covered spaces in the reviewed cities appears to depend mostly on the implementation stage (\emph{is this only a pilot study, as in Cologne-Nippes or St Pete Beach, or a full-scale deployment, like in Penang?}) and the technology that is used to monitor parking occupation. In this respect, apart from the case of Penang, the most extensive coverages are achieved when mobile cameras (typically aboard ANPR vehicles) are deployed. That being said, the comparison is unfair, insofar as this deployment does not provide truly real-time information, in contrast to static cameras or ground sensors. 

\subsection{Technological implementation}

\begin{figure}[h]
    \footnotesize
    \centering
    \includegraphics[width=0.99\linewidth]{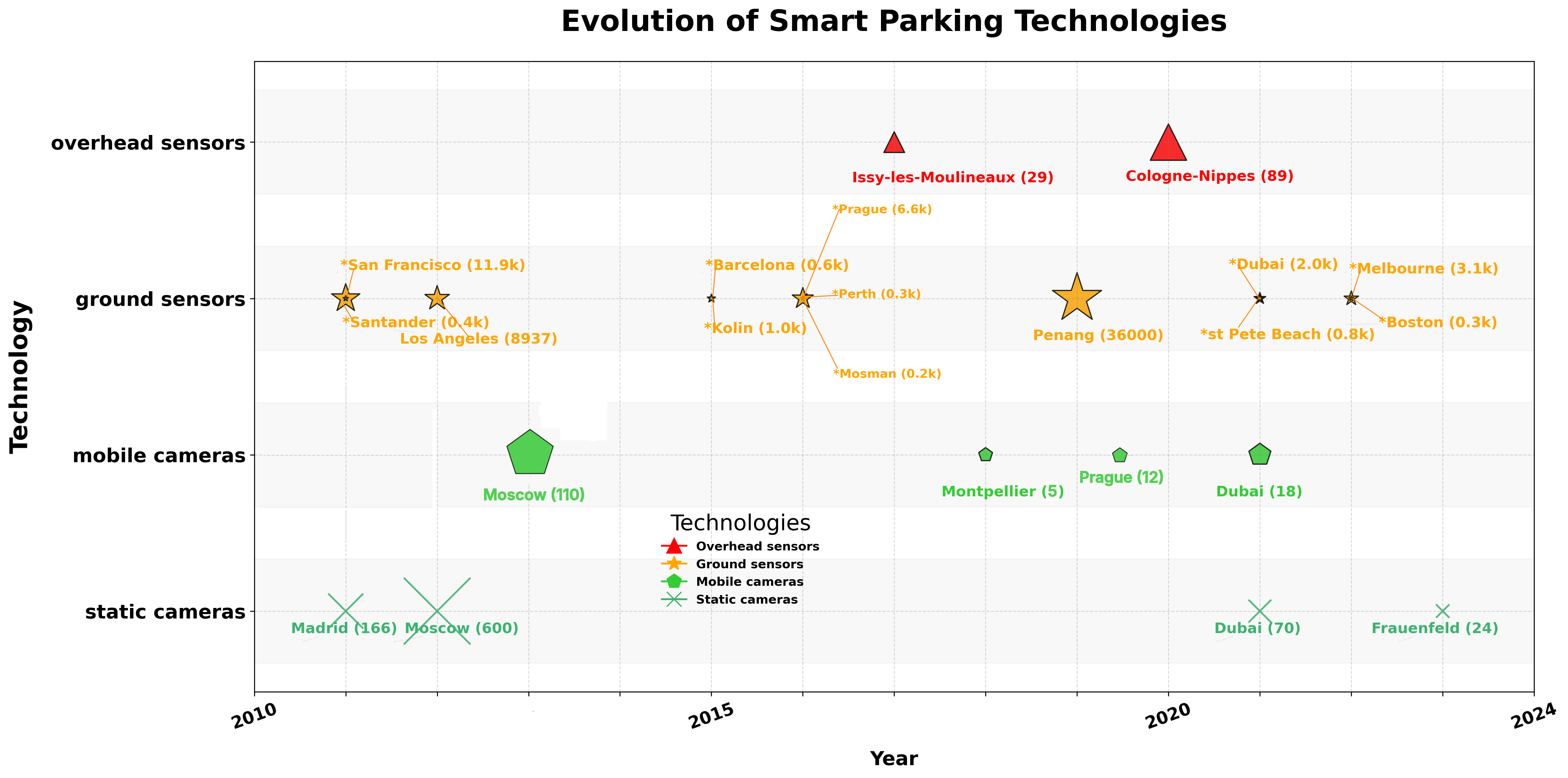}
    \caption{On-street SPS infrastructure. The dates correspond to the beginning of the implementation, whereas the numbers of sensors are updated to take into account project extensions. Note that many more cities are actually using ANPR cars for parking control, but not necessarily promoting any SP functionality beyond control.}
    \label{infrastructure}
\end{figure}

The distribution of sensor and camera technologies used to monitor parking is presented in Fig.~\ref{infrastructure}.
Despite the challenges associated with outdoor detection, diverse solutions have been implemented; we group them into overhead sensors, ground sensors, mobile cameras, and static cameras, regardless of the nature of the waves used by the ground sensors or whether the mobile cameras are embarked in dedicated vehicles or in public buses (the case studies will provide more detail in this regard). 

The number of sensors strongly depends on the strategy, from $10\sim 100$ mobile cameras to $1,000\sim 40,000$ ground sensors. These variations clearly mirror operational differences: ground sensors typically monitor one parking space, but provide real-time information (which may be directly useful for drivers in search of parking), whereas each mobile (embarked) cameras combined with an ANPR or object detection software can cover an extensive number of spaces, but at a lower frequency (less than one control per hour, typically), which may be sufficient to control payment and, if need be, issue citations, thus generating revenue, but not \emph{directly}
relevant for parking search\footnote{Nonetheless, several start-ups leverage AI to make parking occupancy predictions (at diverse time scales) based on these infrequent data.}
. Static cameras cover an intermediate number of parking spaces (from only a few up to several tens), but can be fully unmanned, unlike present-day ANPR cars.

Delving into the details, one realises that the performed categorisation masks a great level of complexity. First, different technologies may be used by the same city. For example, in Dubai as well as in Moscow, ground sensors (in particular, the SENSIT sensors manufactured by NEDAP), static and embarked cameras have all been implemented over the years. Secondly, in the same vein, as we will detail in the case studies, within a given project, the choice of sensors may change over the course of the project, notably favouring cameras over ground sensors at the end of the day in San Francisco, Prague, and Soissons (France). Finally, each hardware category gathers distinct solutions made by various manufacturers and with different properties, not to mention that in several cases SP is not implemented by one main private partner, but shared between multiple partners for the hardware, the deployment on the field, the digital applications, etc. 

\subsection{Information communicated to drivers}

As SP may mainly target  the control of payment (to increase parking revenue) or the continuous collection of data by the transport authority (to adapt parking policy accordingly, e.g., by adjusting parking rates as in SFPark), an SP implementation does not always entail the communication of real-time information about parking availability to the drivers. Even when some information is communicated, it may lack detail or spatial granularity, especially if the goal is `educational', that is, to point to saturation in some areas or at some times and therefore discourage cruising for an on-street spot (see the case studies of Saint Pete Beach and Grand Paris Seine Ouest in Chap.~\ref{casestudiesChap}). 

In the remaining cases, detailed real-time information should be accessible to drivers and guide them to available spots, generally via a smartphone application. However, we have observed that these applications are rarely stable over several years. Instead, they tend to be either plainly discontinued (notably when the company that produced them goes bankrupt), or replaced by other applications (not necessarily sponsored by the city), or integrated into a broader-ranging application (for instance, the parking app of Mosman, Australia, was replaced by a  `Park'nPay' app for all New South Wales). Thus, even when live feeds giving raw parking availability data in real time can be found, e.g., for Melbourne or Los Angeles, there is not always one `go-to' smartphone application that exploits these data and is known to most drivers; this echoes the problem of low public awareness that Peng et al. pointed to in the case of London \cite{peng2017impacts}. The consolidation of the SP market that seems to be taking place may improve the situation in the coming years.

Instructive though they were, the above paragraphs also show the limitations of an objective systematic review as far as SP is concerned. It fails to transcribe the specific context in each city, the actual goals primarily pursued by the stakeholders who decided to implement an SP strategy, and the particularities of the technological solution that was implemented. Also missing is a candid, unpolished appraisal of the effectiveness of the implementation. Therefore, we now move on to case studies for which we interviewed stakeholders.

\section{Case studies}\label{casestudiesChap}

This section gives a concise but all-rounding synthesis of the case studies that were conducted and the feedback from the stakeholders involved in these SP implementations, listed in Tab.~\ref{tab:interviewees}.

\begin{table}
\centering
\begin{tabular}{|p{50mm}|p{40mm}|p{65mm}|}
\hline 
Name & City & Occupation in relation with SP\tabularnewline
\hline 
\hline 
Kenya WHEELER & San Francisco, USA & Main Transportation Planner\tabularnewline
\hline 
Michelle GONZALEZ & St Pete Beach, USA & Director of Transportation and Parking\tabularnewline
\hline 
Nurbayah Iswanis Binti ISHAK & Penang Island, Malaysia & Deputy Accountant \tabularnewline
\hline 
Caroline SACK KENDEM & Douala, Cameroon & CEO of \emph{EZ Park} \tabularnewline
\hline 
 Kevin BOULANGER & Soissons, France & IT Manager\tabularnewline
\hline 
 Hugo LAMBERT & Grand Paris Seine Ouest, France & Deputy Director for Mobility \tabularnewline
\hline 
Peter KOCH \newline
Andrea FOSSATI
 & Frauenfeld, Switzerland & 
Director of the Communication and Economy Service \newline
CEO of Parquery AG
\tabularnewline
\hline 
Georg SPYCHER & Zürich, Switzerland & General Manager at Parking Zürich AG\tabularnewline
\hline 
Justin McKIRDY & Perth, Australia & Executive Director Urban Mobility with Department of Transport and Major Infrastructure (DTMI)\tabularnewline
\hline 

\end{tabular}
\caption{
Identity and position of the stakeholders that were interviewed for the present paper.
\label{tab:interviewees}
}
\end{table}

\subsection{San Francisco, California, United States}

\subsubsection{City Context and Launch of \emph{SFPark}}
San Francisco  lies in the Bay Area in the North of California. With a municipal population slightly below 900,000 residents and a density above 7,000 inh/km$^2$, it is the second densest city in the United States.

The city features different types of parking spaces, some of which are colour-painted on the curb, e.g., in green for short-term parking (less than 15 or 30 minutes). In residential permit areas, cars can generally park for up to two hours in the same space with no permit. In any event, cars should not stay parked for more than 3 days. Parking meters, when present, indicate paid on-street parking and one can pay for durations up to 2 (or 4) hours, either cash or via a mobile application. Even though on-street parking spaces contribute an estimated 63\% to the total parking supply, there is a significant off-street parking capacity, but at rates that are generally deemed quite expensive (more than US\$$\,7\sim 10$/hour) compared to the on-street alternative (typically, from US\$$\,1$ to US\$$\,7$ per hour).

The city is widely known for the difficulty to find parking spaces. That being said, this scarcity of available spaces does not necessarily generate significant cruising traffic or long search times. Indeed, prominent researchers have argued that it is well assimilated by drivers, who thus tend to park some distance away (and ahead) of the busiest areas, so that cruising remains limited \cite{millard2020parking}.

In 2010$\sim$2011, San Francisco pioneered the large-scale deployment of an SP system with a pilot study dubbed \emph{SFPark} and funded by a Federal Highway Administration grant. Ground sensors were installed in 8,200 parking spots, out of the 28,000 metered spaces in the city, with the intent to improve real-time occupancy monitoring and inform demand-responsive pricing: between 2011 and 2014, parking rates in the considered neighbourhoods were periodically adjusted to meet the occupancy target range, i.e., 60\% to 80\% occupancy. The availability and prices were displayed on a dedicated website and mobile application. The pilot was terminated in 2014.

\subsubsection{Interview with Mr. Kenya WHEELER, Principal Transportation Planner, San Francisco}
Due to its pioneering nature and to the design of the project, and unlike the other case studies in the present paper, \emph{SFPark} was thoroughly evaluated, notably in a report published in June 2014 \cite{SFPark2014evaluation}. The report states that the pilot was successful in many respects, with a target block occupancy that was met 31\% more often after implementation, with reported search times that decreased by 43\%, and with an estimated 30\% reduction of greenhouse gas emissions during parking search \cite{SFPark2014evaluation}.
Notwithstanding this detailed evaluation, we were eager to interview an operational stakeholder, Mr. Kenya WHEELER, Principal Transportation Planner at San Francisco Transport Authority, with the hope that the oral format of the discussion would supplement published results with candid opinions and the  $10^{+}$ years that have elapsed would bring additional hindsight. Indeed, some of the feedback reported below stresses points that did not particularly catch the eye previously.

\paragraph{Sensors} Wireless magnetometer-based sensors designed by StreetSmart Technologies were buried in the ground in each parking space in select neighbourhoods; they detected whenever vehicles enter and exit a parking space \cite{SFPark2014sensors}. For the \emph{SFPark} pilot, sensor quality standards had been fixed, but Mr. WHEELER insists that such sensors were a ``very nascent technology'' at that time and San Francisco was the first city to deploy them on a large scale. Still, he deems that in the first phase sensors worked well and the standards were generally met, with an accuracy of 86\% to 90\%. This was ascertained by validation on the ground and computer vision (as is currently tested in Seattle). The Interviewee adds that, shortly before our meeting, he had learnt that a comparison between in-ground sensors, above-ground sensors and cameras had shown a good match, even though the different technologies ``tend to pick different events''. Nonetheless, during \emph{SFPark}, certain areas and factors caused major difficulties. In particular, close to train tracks, electromagnetic interferences due to the passage of trains, streetcars, and even buses interfered with the magnetometers. This led the transport authority to eventually ``pivot to another vendor'' and ``plac[e] two sensors in nearly half of all parking spaces'' \cite{SFPark2014evaluation}.
Still, the data were marred by misdetections and false positives, and they were not reliable enough to calculate precise turnover rates and parking durations; the occupancy was aggregated in time for more reliability.\\

Above all, the sensor performance ``degraded over time, [with a] big drop mid-2013'', a bit more than one year after their installation. In December 2013, ``the vast majority of sensors were off-line''. It seems that one of the major issues was the durability of the hardware; the promised lifespan of 5 to 10 years clearly failed to be met.  As they were self-contained boxes buried in the asphalt, it was not possible to just replace the battery. Thus, they were active for about 3 years, but then there was no funding to replace them.

\paragraph{Connectivity} A standard of  85\% had been set for the percentage of events received within 60 seconds. This connectivity requirement was also found to be a practically relevant challenge, especially in a ``hilly city'' like San Francisco and relying on cell phone carriers and the 2G-3G network.

\begin{figure}[h]
    \begin{center}
  \includegraphics[width=0.5\textwidth]{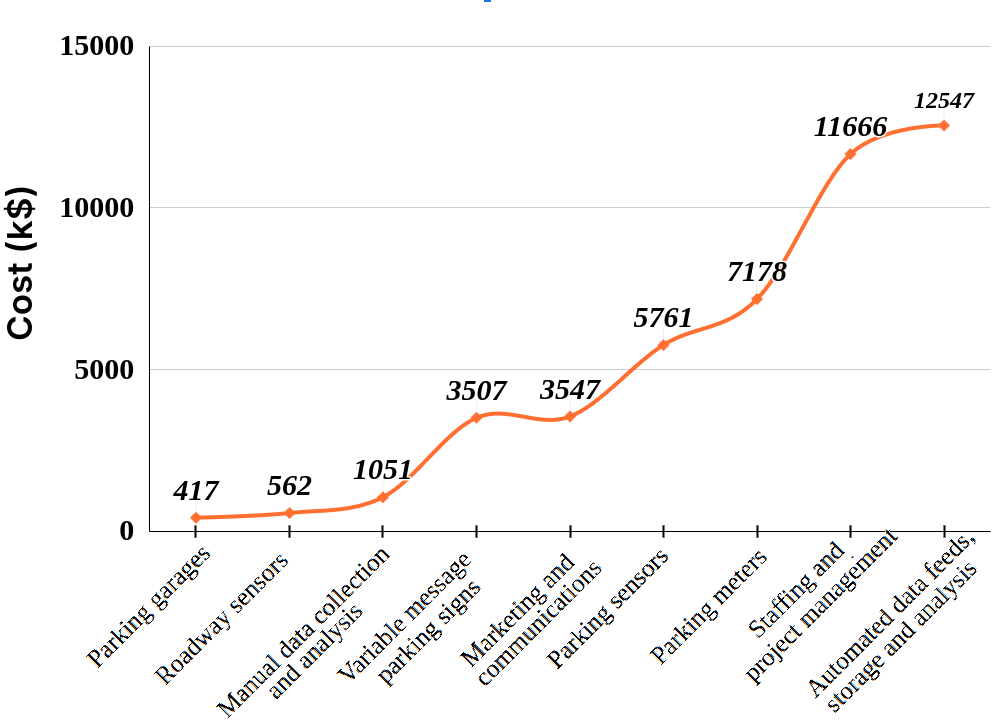}
  \end{center}
    \caption{Cost structure of the 2011-2014 \emph{SFPark} project.\\
    Source of the data: \cite{sfp_pilottheory,sfp_evaluation}.}
    \label{sfpcost}
\end{figure}

\paragraph{Financial aspects}
Mr. WHEELER insists that such innovative projects have only been possible with the financial support of (federal) grants -- from the  Federal Highway Administration in the case of \emph{SFPark}; once the pilot was over, the ``costs [to maintain and replace the sensors] could not be covered without new external funding'' and their operation was thus discontinued.
Figure \ref{sfpcost} shows the cost breakdown of the solution between 2011 and 2014.\\

In terms of revenue, the project generated measurable benefits by making it easier to pay, so that more drivers paid the meter and compliance rates rose, possibly ``increasing public space revenue by \$6.5M''; that being said, the observed reduction in citation revenue somewhat tempers this benefit. \\

The foregoing points highlight a structural challenge: while smart infrastructure pilots may prove effective technically, scaling them requires long-term capital planning.

\paragraph{Impact on traffic} \emph{SFPark} belongs to the few SP projects whose quantitative impact on traffic and on the environment was rigorously assessed. Given the downturn in the economy during the project, which curbed down traffic, it is hard to compare the \emph{before} and \emph{after} situations. In fact, the most compelling analysis probably stems from the comparison of areas affected by the pilot and control areas. Search times were estimated by surveying drivers and by collecting floating car data with staff driving along pre-determined routes and measuring their parking search times. The reported search times decreased from 11.6 minutes to 6.6 minutes on average, which marks a 43\% reduction in pilot areas, as compared to the 13\% reduction in control areas \cite{SFPark2014evaluation}. This aligns well with the observation that block-faces were full a much smaller fraction of the time in pilot areas, whereas (perhaps surprisingly) they tended to be full more often in control areas. \\
Greenhouse gas emissions were estimated based on these data and the resulting savings in vehicle miles traveled per day (in contrast with more recent smart grant projects, e.g., in Minneapolis, where gas ``sensors have been deployed at intersections''). It was found that  emissions decreased by 30\% in pilot areas vs. 6\% in control areas.

\paragraph{Perspectives and personal opinion}
The Respondent underlines that, despite the technical issues encountered in this pioneering project, it generated a wealth of data that are being used up to this day.  
 For instance, the results were used to implement city-wide rate adjustments depending on the demand in 2017$\sim$2018. The lessons remain highly instructive for cities weighing the trade-offs between high-cost, maintenance-intensive hardware solutions and lower-cost, policy-driven approaches such as dynamic meter pricing and enforcement.\\

Touching on perspectives in terms of policy, the Interviewee emphasised the dependency on grants to implement new projects. Recently, Seattle and Minneapolis got funding for their project ``to price the curb with automated tools''; Mr. WHEELER hopes that ``they will move forward and inform other cities''. Otherwise, he admits that the day-to-day operations by the Transportation Department are mostly \emph{ad hoc} because of limited resources and driven by complaints from residents and merchants. Still, the department is pro-active in promoting streetscape projects and proposals to shift parking zones.\\

Today's challenges partly mirror those of 2011, insofar as there is a finite parking supply and a tremendous demand. Much boils down to ``getting vehicles to follow city regulations'': often the ``curb activity does not match regulation'' and it is frequently used for drop-off or illegal parking (e.g., not in the right direction) which may force cars on the road to change lanes or ``even block a train, delaying hundreds of people''. An idea could be to set ``cameras on trains to take photographs and issue citations''. In the past few years, the uptick in e-commerce, the upscaling of ride-hailing companies (raising a ``huge challenge'' for curb management), and the spreading of autonomous vehicles have been points of concern. Often, the City cannot regulate on its own, as the prerogative is held by the State. In this respect, says the Respondent, the pandemic emergency softened the restrictions imposed on the transportation department activities  and shortened decisional circuit, thus facilitating changes. More generally, it is usual that technology takes precedence over regulation, making the tension between the will to ``let technology go forward'' and the need for regulation palpable. Similar tensions are also felt strongly when it comes to  encouraging softer modes of transportation by means of changes that go counter to particular habits or interests.

\subsection{St Pete Beach, Florida, United States}

\subsubsection{City Context and Launch of the Smart Parking Project}
The City of St. Pete Beach, a coastal community in Pinellas County, Florida (USA), with a population of about 10,000 resident and a much larger seasonal visitor base, piloted an on-street smart parking program in 2021 to alleviate the heavy traffic and in particular address the challenges of beachside parking demand. Under unchanged parking supply, i.e., with around 1,300 (diagonal and parallel) beach parking spaces, including both permitted and paid parking, the program revolved around the installation on the metered spaces of approximately 750 radar ground sensors by the company \emph{Libelium}, transmitting real-time availability data to a cloud operated by \emph{Conure}, which was displayed on the City's website and on a \emph{ParkMobile} app, also used for payment. The initial pilot was launched for free on 9th Avenue, before expanding in May 2021 to all parking areas in the southernmost district of Pass-a-Grille (one of the city’s busiest waterfront districts, for which there is mainly one way in and one way out, with a long-stretch street along the beach). In parallel, the existing free on-demand electric shuttle (called \emph{Freebee}) expanded its rides, even though public transit to the southern beaches remains scarce, partly due to the residents' objections to having buses on the road.  While plans for a SP roll-out in the whole city were envisioned, in early 2023, the City Commissioner  decided to pause the program.

\begin{figure}[h]
    \footnotesize
    \centering
    \begin{overpic}[width = 0.7\linewidth]{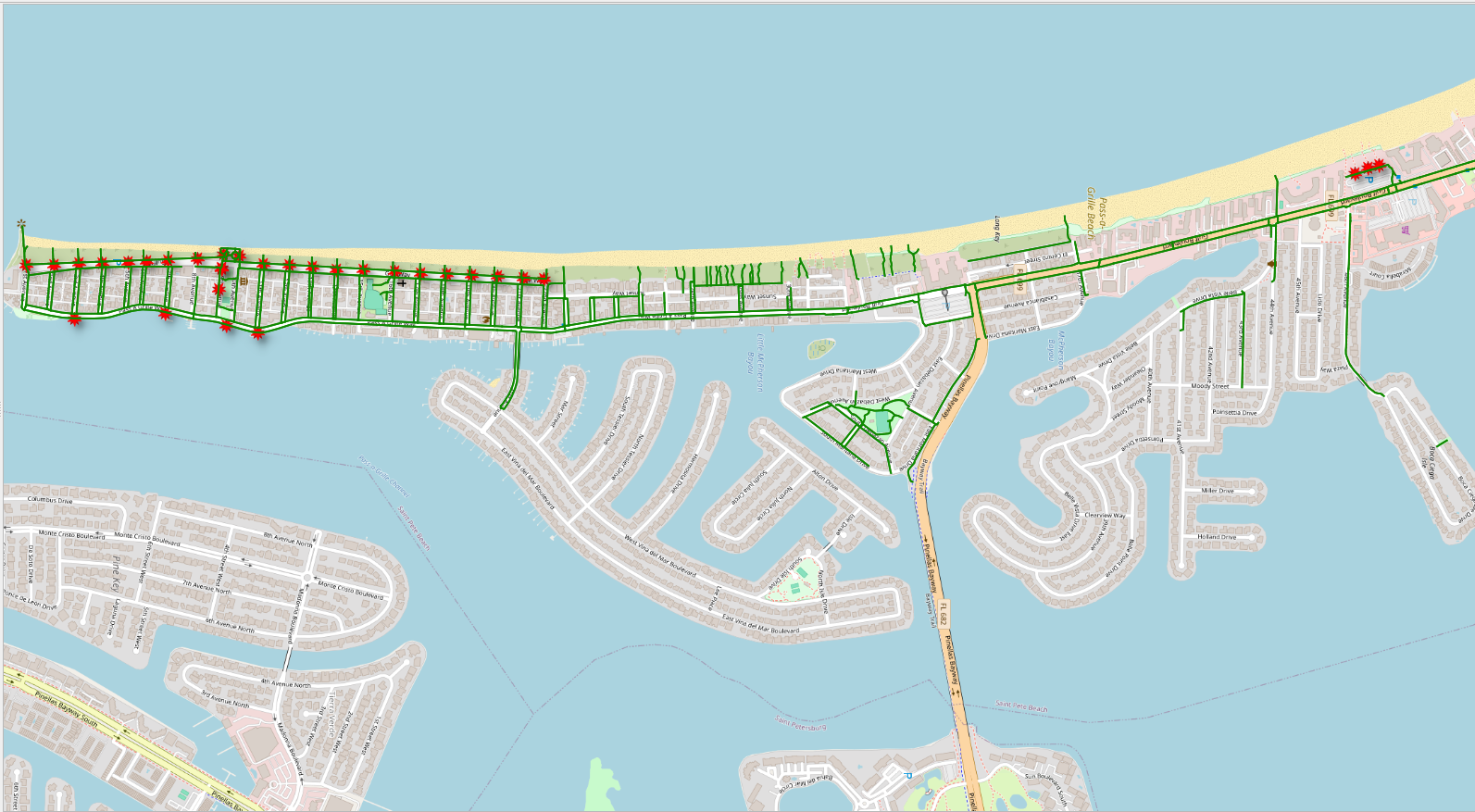}
    \put(45,0){\includegraphics[width = 0.3\linewidth]{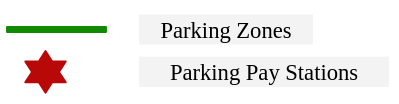}}
    \end{overpic}
    \caption{Parking Zones and Pay Stations in St Pete Beach. \\
    Source of the geolocalised data:\cite{ArcGISWebApp_CSPB_2025}, OpenStreetMap. The figure was produced by the authors. }
    \label{stpeteebeach}
\end{figure}

Parking rates were \$4/h from Monday to Friday;  \$4.75/h over the weekends and a \$25/day flat rate for parking lots at boat ramps and on public holidays. Payments are made through the \emph{ParkMobile} app or through the 36 parking meters shown in Fig.~\ref{stpeteebeach}. Residents have access to parking permits. Lastly, note that two-hour free public parking is provided along 8th Avenue, to promote shopping. 

\subsubsection{Insights from Michelle GONZALEZ, former Transportation and Parking Director for the City of St Pete Beach, USA}

\paragraph{Main purpose and evolution of the project}
Parking was widely regarded as an issue. But our Interviewee, Ms GONZALEZ, explains that one of the main initial goals was actually to divert traffic away from the beach-side road by showing drivers the lack of vacant spaces and urging them to park elsewhere (or even choose another beach). Indeed, the configuration of the beach-side road implies that drivers entering it will not miss a free spot along it.
Communication and marketing efforts were made to educate the public to the existence of this solution. The program was suspended and the ground sensors in 2023 for `political reasons' and `aesthetic issues', notably to restore the historical character of the place.

\paragraph{Sensor technology}
Residents' concerns played a significant role in the choice of SP hardware and its appraisal. In particular, cameras were deliberately avoided because of privacy concerns on their part: despite their potential, ANPR and cameras ``might not have been a very popular tool to use'' in that community.   After the sensors (which are thin disks on the ground) were installed, some residents expressed dissatisfaction for aesthetic reasons, because they might not fit a historic place. \\
In terms of performance, ``the sensors initially performed well, but issues began to arise around November 2022. At that point, the program had been running for just over a year. We worked with the vendor to replace and recalibrate the sensors,'' Ms GONZALEZ said, pointing to the need of yearly routine maintenance. At that point, before maintenance, 25\% of the sensors possibly exhibited malfunctions.

\paragraph{Financial perspective}
Financing the project was ``not an issue'' for the city. The pilot was offered mostly for free by the vendor and then the direct implementation cost amounted to \$25,000/year, with a prospective full SP cost of order \$100,000/year (tentative figure), to be compared with an annual parking revenue around \$4,000,000. (Now that the project is over, excess parking revenue finances a hurricane mitigation fund). SP costs would possibly have been outweighed by additional parking revenue. Parking revenue grew over the course of the project, but this was only coincidental and due to a rise in parking rates rose during,  Ms GONZALEZ explains. "The solution was not centered around revenue generation''.

\paragraph{Impact on traffic} There were qualitative observations of heavy traffic prior to the implementation of the project, but no quantitative diagnosis. The project was halted before it reached its final expansion stage and its impact on traffic was never assessed quantitatively. There was no report of a visible impact on traffic.

\paragraph{Users' satisfaction} According to the Interviewee, a lot of positive feedback was received from users (mostly local drivers) in the first stage. However, after about one year, concomitantly with the emerging sensor deficiencies, some users reported inaccuracies. One of the causes for the complaints about inaccuracies, Ms. GONZALEZ says, is that a parking spot shown as vacant could get busy while a driver was heading for it; given that the road is quite busy with drivers looking for parking, being outrun by another driver may have been a common problem. On the other hand, hardware deficiencies were also observed by the staff on the field. In any event, this eroded drivers' confidence, generating complaints and undermining public acceptance. \\
At a broader scale (not specifically for St Pete Beach), \hyperref[https://parkmobile.app.link/SYLSYDYShR]{ParkMobile} users express satisfaction with the app with few concerns around Payment processing problems, interface, stability, low customer support, account security, and device compatibility.

\paragraph{Actual benefits}
The City previously relied on  manual operations (including chalking of tires) by 5 parking enforcement officers (3 active on a daily basis) to ensure compliance; about 4,000 citations were issued each year. The SP system enabled City officers to work more efficiently and directed them to relevant spaces, so that, with the same staff, the number of citations a year was tripled. At the same time, SP also probably increased compliance.

\paragraph{Personal opinion on perspectives} Ms. GONZALEZ thinks that SP would have had a larger impact if it had been deployed city-wide, so that people can make a fully informed decision as to where to drive and try to park. The idea of installing a variable message sign (VMS) indicating if parking was saturated or not had been going around. Parking reservation could also have been considered in the future, but not in the first stage, and it would not have been a solution to all problems. Overall, she recommends choosing solutions that are customisable, flexible and integrable; in particular, collecting all relevant information in the same database is a non-negligible asset. \\  From a broader perspective, Ms. GONZALEZ insists that  SP initiatives should strongly depend on the location one considers, even in Florida. In big metropolises like Miami, above all one should give people options on how they want to travel and promote transit. In smaller cities with poorer public transit possibilities, the situation is different. ``There is no one-size-fits-all'' solution.




\subsection{Penang Island, Malaysia}

\subsubsection{City Context and Launch of the Smart Parking Project}

Penang, with a population of nearly 1.8 million inhabitants, is a State of Malaysia which spreads across Penang Island and the mainland.
Travelling by car is largely ingrained in local people's habits and parking demand far exceeds supply, especially in the city of George Town (on the Eastern coast of the island), where public parking is generally saturated. Creations of more parking spaces to remedy their shortage are strongly limited by the land space that can be used for this purpose.
Parking options consist of public (Government-owned and managed) parking spaces on the curb, and exclusively private off-street parking. The former are much cheaper, with typical costs around 0.8 US\$ per hour, as of early 2025, as compared to US\$3 for private parking. Therefore, drivers prefer parking on-street and are often ready to cruise for public parking. This is particularly true in the centre of George Town, where people may commute for work, to go shopping, and available public parking is generally very scarce.

As part of \hyperref[https://penang2030.com/]{Penang 2030 vision}, the Penang Island and Seberang Perai City Councils (MBPP \& MBSP) launched an SP initiative in August 2019 to manage the constrained urban space, teaming up with HeiTech Padu Berhad for the development of the system, Sigfox for sensor transmission, and Xperanti for IoT 0G network. As of February 2025, the system covered 37,000 parking spots, about 24,000 on the mainland and 13,000 on the island, using surface-mounted (ground) sensors for detection, powered by long-life batteries that are suppsed to last up to ten years, and license plate recognition for enforcement. Real-time information is communicated to a central dashboard used by City officers and it is also made available to drivers via the \emph{Penang Smart Parking} mobile application, which enables users to locate and navigate to vacant spots, pay the parking fee, and possibly extend it. According to the official figures communicated to us, the application has been downloaded over 1.5 million times  as of early 2025 and there were 293,000 active users in January 2025. (Incidentally, the year after the implementation, i.e., in 2020, a study based on a survey of around 600 people had arrived at the conclusion that only a third of residents were aware of the SP application and that not enough publicity had been made about it \cite{kee2020assessment}.)

\subsubsection{Insights from Ms. Nurbayah Iswanis Binti Ishak, Deputy Accountant at Penang Island City Council}

\paragraph{Main purpose and evolution of the project}
The Interviewee, Deputy Accountant at Penang Island City Council, thinks that it is one of the governmental missions of the City Councils to provide people with affordable options to travel by car, in parallel to promoting transit and walking. In particular, in light of the car-centered habits of Penang citizens and the restricted land available to construct more surface parking, SP appears as a service that enhances drivers' experience, despite the aforementioned constraints.  \\
Furthermore, as part of the Clean and Green initiative, SP is also aimed at ``reducing paper use'' (notably for citations) and ``reducing cruising'' by indicating where there are vacant spots, or redirecting drivers to private parking (knowing that drivers tend to search for the cheaper option of parking on the curb in the first place).

\paragraph{Sensor technology}
The City and its providers opted for ground sensors, which are claimed to take less than 30 minutes to mount on each spot, and managed to equip most of the parking spaces in George Town (12,000 sensors initially, then increased to 13,000, for 18,000 on-street spaces) as well as on the mainland (24,000 sensors). \\
Overall, the sensors are perceived as working well, with allegedly more than 90\% accuracy. Nonetheless, it may happen that the staff on the field observers that sensors malfunction or break, in which case the inaccuracies are reported to the vendor. Sensors tend to be ``overpredict'' the occupation, that is to stay that spaces are reported as occupied when they are actually vacant.

\paragraph{Financial perspective}
Thanks to a private finance initiative, the City Council did not have to make any substantial investment for the implementation of the system. Instead, the agreement with the system developer relied on sharing the parking revenue (details of the sharing agreement are confidential).

In terms of financial added value, the SP system provided an efficient way to track revenue in real-time and to help enforcement officers on the ground control parking compliance.
Besides, an increase of the public parking rate to US\$1.6 was due for March 2025, generating some frustration in the population, but deemed necessary to maintain service quality (possibly including sensor maintenance) and security. However, it should be underlined that public on-street parking remains considerably cheaper than private off-street alternatives, as providing affordable public parking is regarded as a governmental mission; accordingly, it is not the goal of the City Council to maximise parking revenue, which ``is not the main (source of) revenue''.

That being said, significant \emph{losses} in revenue occur whenever the SP system is down and these are considered a major challenge by the Interviewee. In these situations, due to the centralised nature of the system, both payment and parking enforcement are affected and, as a result, drivers are informed on social media, notably, that they can park freely. Such breakdowns are reported to the vendor and happen ``at least once a year''.

\paragraph{Users' satisfaction} 
Among complaints from drivers, the Deputy Accountant lists frustration that there is not more parking space (which she says cannot be remedied because ``we do not have much land in Penang, we cannot expand our parking infrastructure'') and that public parking fees are planned to increase. Regarding the SP smartphone application (\emph{Penang Smart Parking}), it is largely used. While some users express concerns about missing functionalities, the unavailability of the service from time to time, compatibility and account security problems,  others appreciate its efficiency.\footnote{We cannot comment on how user-friendly the application actually is, because we managed to download and install the app without problems, but cannot use it from abroad.}

\paragraph{Impact on traffic} Some impressive figures are mentioned on corporate websites about the impact of SP on the cruising traffic in Penang, e.g., 50 percent reduction in the time needed for users to find parking and 30 percent reduction in  $\mathrm{CO}_2$ emissions, but their origin cannot be sourced, which casts doubt on their reliability. The Interviewee indicates that these figures do not come from the Department. Nonetheless, since SP helps drivers locate unoccupied parking spaces throughout Penang or redirect to a private parking option in case no public space is available, she is convinced that qualitatively SP has had a positive effect on  traffic.

\paragraph{Actual benefits} The introduction of SP has facilitated parking enforcement in a cost-effective way, without public capital expenditure, and it has ``reduced paper use''.

\paragraph{Personal opinion on perspectives} 
The Interviewee agrees that making parking easier cannot be the be-all and end-all of transportation policy and that the City Council has to promote walking and usage public transportation. However, she thinks that these goals should be conducted in parallel to the deployment of SP and that they are not directly in conflict, because presently citizens prefer using private cars and public parking is saturated in any event. (For this reason, she does not expect the planned hike in parking rate to significantly affect parking tension.)

\subsection{Douala, Cameroon}
\subsubsection{City Context}
With a population around 4.5 million inhabitants, Douala is the largest city in Cameroon and its economic capital, notably thanks to its harbour. Its population tripled between 1980 and 2000 and keeps growing at a brisk pace. The city routinely suffers from congestion and many complain about the lack of parking spaces, especially in the business districts, and the chaotic parking situation.

On-street parking is charged at a rate 100 CFA Francs (around €0.15) per hour, or 500 CFA Francs for the whole day; this amount can be paid directly to municipal agents on the ground. Note that, as of 2024, the city was looking for a service provider in charge of parking enforcement. The numbering and marking on the ground of parking spaces is quite recent. A pilot study for smart on-street parking, involving online payment and real-time occupancy monitoring on about 200 parking spaces, was started in 2021 with the company \emph{EZ Park}. In 2025, the company put to test a more autonomous solution.

\subsubsection{Insights from Ms. Caroline SACK KENDEM, CEO of \emph{EZ Park} }
Caroline SACK KENDEM, who founded the \emph{EZ Park} start-up in 2019 and now heads this company, answered our questions. 

\paragraph{Motivations}
An ``entrepreneur at heart'', Ms. SACK KENDEM realised that congestion was becoming a paramount issue in several African metropolises, including Douala, and was eager to provide an IoT solution to the more and more acute parking problem. She has observed that, when they need to go to a place with high parking tension, residents often ``choose to be chauffeured'' there and let the chauffeur in the car while they run errands.  For longer stays, people need to park in peripheral streets or on the pavement, no private parking is supplied by e.g. their firm. The Interviewee set to offer a solution that would be operational not only for Douala, Cameroon, but all over Africa. 
Beyond Douala, she also prospects the market in Ivory Coast, Nigeria, Zambia, and Rwanda. She targeted on-street SP because the ``technological barrier is higher'' there, which makes it harder for competitors to enter the market.

\paragraph{Implementation of the service}
The first version of the service was developed with the technological collaboration of a French company and, after all technical difficulties had been solved, it was put to the test as a pilot study on a parking zone with 200$\sim$300 dedicated and marked spaces in Douala from November 2021 to April 2022. This first phase revolved around a mobile application where drivers could register their licence plate and pay for their parking online. The system registered 600 to 700 subscriptions in the period; users were able to pay online, but not to check for vacant spots. A back-office dashboard enabled city officers to monitor payments and centralise field data in real time, including the licence plates of illegally parked cars. To control parking, 10 agents already in charge of parking enforcement were enrolled in the pilot and had to take photographs of the licence plates of parked cars, whereby they could determine which ones had no transaction record in the system. \\
A second, more autonomous SP solution was launched at the end of 2024 and was still under test at the time of our interview. Rather than extensive manual control, this IoT solution relies on sensors buried in the ground and able to detect the presence of a vehicle in real time, but also to communicate with a contactless pass in the parked car and 
operate a transaction. The sensors are equipped with a battery that can last up to three hours in case of power cuts. Installing them ``does not represent a significant financial burden''.

\paragraph{Actual benefits}
A first major benefit for cities, says the Respondent, is that the SP solution markedly increases their parking revenue. While manual parking enforcement was partly ineffective in Douala, she details that  the parking revenue for the zone of interest ``was multiplied by 4'' during the first pilot. 
This is also related to the specifics of parking enforcement by agents, who are typically assigned a \emph{fixed minimum} amount to collect every day; this is an exhausting job, especially in harsh weather conditions, in the heat or with heavy rainfall, and without SP guidance agents do not complete their task in the most efficient way. The digital solution ``optimises their operations'' and reduces the tensions with drivers. Finally, the cashless solution is a step forward, insofar as drivers can no longer ``argue that they have no cash'' in order to escape payment and park for free. Nevertheless, Ms.  SACK KENDEM claims that users were very satisfied with the convenience of this mobile solution.

\paragraph{Difficulties}
Many difficulties mentioned by the Interviewee are related to financial disagreements with municipal services, particularly in Cameroon, and conflicts between political and economic interests. For instance, all the cities she engaged with are unwilling to dismiss their parking enforcement agents; a less human-power-intensive control solution thus requires assigning these agents different tasks. Besides, the solution needs to be approved by different stakeholders in the municipal services and they can have conflicting views.\\

There was also major disagreement as to when the revenue should be shared, i.e., as soon as a parking payment is made, following the entrepreneur's desire, or at the end of the month, as imposed by the Douala municipality, while city officials in Ivory Coast are more flexible in this regard. Finally, the parking rate is too low. While it was set at 100 CFA Francs (€0.15) at the time of the interview, she argues that it should be raised to 300 or 500 Francs for the solution to be profitable. This rise would also urge people to park farther away, on cheaper parking spots, and then use a dedicated shuttle or bus to their destination, an idea shared by municipalities.

\paragraph{Impact on traffic}
 Ms.  SACK KENDEM is not sure that the pilot had an impact on congestion, but she thinks that drivers heading for the pilot parking zone spent less time searching for parking.

\paragraph{Personal feedback and perspectives}
Regarding the first pilot, the Interviewee highlights that massive communication on the radio and on TV was needed beforehand, in addition to  duly marking (i.e., painting) the involved parking spaces. As a result of these efforts, she believes that virtually all local drivers were aware of the digital solution.\\
She does not plan to extend the SP solution to the whole city, as she considers that it should be restricted to the zones with maximal traffic and parking tension. \\
Eventually, she notes that some cities (e.g., in Ivory Coast) have shown a clear interest for an on-street parking reservation option. She believes that this is practically doable, but for an increased rate and only for a fraction (10\%, 15\%?) of the equipped parking supply. This option would not necessarily urge people to drive more because the ``parking rate would be an obstacle''.

\subsection{Montpellier, France}

\subsubsection{City Context and Launch of the On-Street Parking Project}
Montpellier is a city of 310,000 inhabitants (615,000 if the whole urban area is included) located in the South of France, close to the Mediterranean coast. Its public transportation network consists of 5 tramway lines and 41 bus lines. In the city centre, there is a sizable pedestrian area. The city is known for its pro-active transportation policy and in December 2023 became the first French metropolis to make public transportation free for residents.

In parallel, while the street network was redesigned in 2022, adding up to the constraints faced by motorists, Montpellier insisted that transport modes should not be regarded as conflictual and invested in the development of SP and of a ``parking observatory'', and the design of a smart parking guidance algorithm. 
The urban core is served by about 20 off-street parking facilities and over 20,000 on-street parking spaces, for commerce, transit, tourism, etc. There are four distinct paid parking zones, totalling 18,000 spaces: a green zone for day-long parking, an orange zone to park for less than 5 hours, a yellow zone for parking durations typically less than 2 hours, and a red zone for short-term parking (typically less than half an hour); parking rates in the first three zones range from 1€ to 2.5€ per hour. Such a differentiation encourages greater turnover in saturated areas while keeping short-term stops affordable. 
To control parking payment and gather valuable spatial data, 3 (and then 5) ANPR cars have been deployed since 2018, replacing manual control by agents on the ground. These vehicles, equipped with high-tech roof-mounted cameras and driven in turn by a pool of 15 staff members, scan up to 10,000 license plates per day. In addition, ground sensors have been installed on spaces reserved for disabled people and parking spaces in the red zone.


\subsubsection{Insights from our collaboration with the transport operator}
At the end of 2023, we initiated a collaboration with \emph{Transports de l'agglomération de Montpellier} (TaM), which manages and operates transport by virtue of a delegation of public service, in order to guide the design of their smart parking algorithm. In the frame of this collaboration, we have conducted field surveys, analysed massive parking data, and interacted with several TaM executives. The following summarises the insights we have gained.

\paragraph{Motivations}  
The fleet of ANPR cars primarily target the control of parking payment. In addition, the Montpellier metropolitan mobility department was eager to leverage the massive data thus collected to gain insight into spatially and temporally resolved parking occupancy, via an online observatory that feeds on the collected data. Finally, a couple of years after the deployment of the system, TaM invested in the development of a mobile-based smart navigation application, which guides drivers to a suitable parking area close to their destination, in light of parking occupancy predictions.

\paragraph{Technology}
Montpellier opted for ANPR-based mobile surveillance, offering a more cost-efficient and scalable solution than ground sensors. The detection and geolocation of licence plates, and the processing and management of the collected data are operated by private companies, \emph{Arvoo} and \emph{Egis}, respectively, as of 2025.
The accuracy of this approach for the detection of parked cars (including closely packed cars) was first demonstrated to TaM in specific street settings. Then, in the framework of our contract with TaM, we quantitatively evaluated this accuracy on a larger scale in 2025 by manually checking the occupancy of parking spaces (clustered in ``polygons'' of 1 to 20 spaces) along the typical trajectory of an ANPR car (with two agents performing their counts independently) and comparing these geolocated manual counts with the simultaneous ANPR detections, extracted from an online database. We found that the ANPR detections were quite accurate, with an error (root-mean square error (RMSE) around 0.85 cars per polygon; absolute error (AE) around 0.5 cars per polygon)  that was only slightly larger than the difference between the manual counts performed by the two agents (RMSE$\simeq 0.65$ cars/polygon; AE$\simeq 0.4$ cars/polygon). While these raw detections were thus quite accurate, technical problems were noticed in the post-processing stage, when the per-polygon occupancy is aggregated spatio-temporally, and remedial action was suggested.

\paragraph{Insights into parking behaviour}
Globally, the occupancy (on a Thursday in June, with no exceptional event) was around $60\%\sim70\%$, but high spatial disparities were observed, with an average occupancy around 70\% in the urban core and around $40\%\sim50\%$ in more peripheral areas, all days of the week (these averages only weakly depend on the weekday between Monday and Saturday). As expected, spaces reserved for delivery or for disabled people exhibit much lower occupancy. Interestingly, we noticed that the massive and finely resolved data collected by TaM also reveal finer features of mobility behaviour, such as slightly higher on-street parking occupancy when the weather was exceptionally cold for the considered month (particularly in April and May) or hot (for most months, except the summer holidays). Besides, more rainfall systematically correlates with higher parking occupancy.

\paragraph{Financial aspects} TaM relies on the ANPR fleet for paid parking control (exclusive of the control of non-paid areas, whether they correspond to free parking or illegal parking). The parking revenue thus heavily depends on the quality of the control by ANPR cars. Nevertheless, Montpellier opted for a fairly permissive policy, in that for each detected mismatch between parked cars and parking transactions a photograph of the situation taken by the ANPR car is carefully controlled (e.g., to see if the driver was still aboard the car, or walking to the parking meter) and uncertainty benefits the driver; more than half of the suspicions of unpaid parking are thus rejected.

Besides, residents and disabled people benefit from specific rates. Parking meters (and the mobile application) deliver free parking tickets to disabled people. Whenever the corresponding licence plates are detected by the ANPR cars, agents are sent on scooters to check the presence of a valid disabled card. Regarding residents, cars displaying residential parking permits represent about half of the occupancy overall, but generate only a small fraction (about 20\%) of the parking revenue. 


\paragraph{Users' satisfaction}
To collect feedback from motorists, we administered a small field survey  on October 23rd, 2024, asking a dozen people who had just parked and passers-by for their impressions. The results were enlightening. Most drivers declared that they had searched for parking. But, when asked how long it had taken them to find their parking space, at least two of the surveyed drivers manifestly overstated their parking search time (or understood parking search in a very loose sense), insofar as they reported search times of 10 and 15 minutes, whereas several vacant spaces could easily be found in the immediate surroundings. This may echo the observation that actual search times might not be as long as often declared \cite{saki2024cruising}. Most drivers indicated that they had parked in the best possible location, very close to their destination (less than 50 meters away), but distances to the destination of $400\,\mathrm{m}$, $500\,\mathrm{m}$, and $1\,\mathrm{km}$ were also reported (which is not impossible, given that some had parked at the periphery of the pedestrian area). Finally, some global impressions were that it is ``hard to park in the city centre; one has to park far away'', ``parking is expensive and there is no space to park; I moved my shop out of the city centre for that reason'' (a shop owner), ``parking is too expensive'', ``with a disabled card, parking is not an issue'', etc., but these statements should probably be nuanced in light of well known response biases.


\paragraph{Impact on cruising traffic}
Traffic information is released on open-data platforms by the mobility department of Montpellier, but no quantitative analysis of the specific impact of SP-related initiatives is foreseen, to our knowledge.

\subsection{Soissons, France}
\subsubsection{City Context and Launch of the Smart Parking Project}
Soissons is a historic city located 100 kilometers to the North East of Paris, in the midst of a rural territory, where mobility largely revolves around private cars. It has a population of less than 30,000 inhabitants, or about 50,000 if one includes the metropolitan area. 

Drivers largely rely on surface parking, both on the curb (268 on-street spaces in the city centre) and in parking lots (in the centre and at its periphery), even though there are also some off-street parking facilities. The first 15 minutes of on-street parking are free in the city centre. For fairly short-term stays beyond that duration, there are ``blue zones'' that were free for stays of less than 1 or 2 hours or had a preferential rate (the policy changed over the years). 
In case a driver fails to pay, they are liable for an increased fee, which altogether brought an alleged revenue of €27,000 to the city in 2024.

An SP system was deployed in 2018. In 2020, about 1,000 parking spaces were monitored, both on the curb (
Rue du Collège,
Rue du Commerce,
Rue Georges Muzart,
Rue Jean de Dormans,
Rue Saint Martin,
Rue Saint Quentin)
and in surface lots (Esplanade du Mail,
Place Dauphine,
Place de l'Evêché,
Place de l'hôtel de Ville,
Place Fernand Marquigny,
Place Mantoue,
Square Bonnenfant), out of 1,700 in the city centre; their live status could be checked on a dedicated smartphone application, and further extension of the coverage was planned. However, a few years ago, the vendor went bankrupt and the project evolved.

\subsubsection{Insights from Mr. Kevin BOULANGER, IT Manager for the city of Soissons}
In Soissons, the SP project was largely supported by the IT Department, both in terms of resources and humanpower. In particular, the position of officer responsible for Smart City projects, officially created in 2016 and operational from 2018 to 2023, was integrated in the same department as IT. Kevin BOULANGER, who manages the Soissons IT Department and has taken the lead of the Smart City project, is our Interviewee.

\paragraph{Main purpose and evolution}
He believes that the main motivation for the SP project was the political will to encourage Smart City initiatives. Despite its small size, the city of Soissons is quite dynamic in this regard. There was also a desire to gain insight into actual behaviours, in order to influence them, and to limit the number of cars parked for an excessive duration (\emph{stuck cars}). As an additional motivation, some frustration had been expressed by some drivers unable to find a parking space, e.g. in the centrally located shopping street \emph{Rue Saint-Martin}, but this  
is largely due to ingrained car-centric habits in this territory and the ``wish to park right away, right in front of the targeted shop, and not $50\,\mathrm{m}$ or $100\,\mathrm{m}$ away''.
\\
The SP project first involved a dedicated smartphone application (called \emph{ParkingMap}) where drivers could 
spot available spaces in real time. However, there were technical problems, including geolocation, data reliability, and the ``need to zoom on the screen to see available spots'' (which cannot be done while driving). Therefore, the application was discontinued when the \emph{ParkingMap} start-up went bankrupt.
Presently, parking data are in the process of being integrated into the broader-scope application of the city (powered by \emph{NeoCity}), which has been downloaded more than 8,000 times overall and can guide the user to a destination if they click on it. 
But the collected data are \emph{already} displayed on a centralised dashboard for city officers, with updates every 5 minutes approximately (the Respondent was not sure about this point) and more reliability than in the previous user application; they are used to get finer insight into citizens' habits and mobility patterns in a privacy-respective way, in order to improve the urban mobility policy. A further goal was to increase the turnover of parked cars and prevent excessively long parking.

\paragraph{Sensors}
Initially, the SP system relied on ground sensors for a good part, notably for curbside parking. But, as the project evolved, the cost-effectiveness of the solutions was more and more taken into account and camera-based sensors (connected with the 3G network) were favoured over ground sensors. Indeed, while ground sensors only cover one spot each,
the overhead cameras installed by \emph{UpCity} typically monitor 5 on-street spaces each and around 15 spaces in parking lots, hence a choice to focus on surface parking lots. It is interesting to note that technical difficulties have led to the exclusion of some areas from the SP system, notably, parking areas with high turnover rates, which are harder to monitor, and areas where trees and vegetation obstruct overhead vision. Overall, close to 150 sensors have been installed and cover shy of 1,500 surface parking spaces, at least 350 of which are on the curb (including 150 free spaces), scattered across the city centre. \\
In terms of accuracy, results with the first system were deemed problematic for multiple reasons, but the new system. provides ``more reliable'' results.  \\


In parallel, the occupation of ``siloed'' off-street parking facilities is monitored by a counter at the gate, which is ``the most reliable solution''. 

\paragraph{Information display}
Live parking information is displayed on six parking guidance boards at the entrance of the city, which show the number of vacant spots in the different parking lots, and six parking guidance boards in the city centre, which point to the closest parking lot and indicate the number of vacant spots in it.

\paragraph{Financial perspective}
Financially, the French State subsidised 50\% of the expenses; the remaining costs have been supported by the IT Department in Soissons. 
Once the city had invested €180,000 to install the sensors, the annual subscription for the SP service amounts to €5,600 only. Another €200,000 was spent on the parking guidance boards.

\paragraph{Impact on traffic and actual benefits} Qualitatively, there is no severe problem of traffic or parking search in the city, either before or after implementation of SP, although some times of the day and areas are busier. Quantitatively, the \emph{ex ante} traffic conditions were not precisely diagnosed. \\
Above all, the ability to monitor parking occupancy and turnover in real time serves qualitative purposes in terms of policy, ultimately aimed at reducing the number of \emph{stuck cars} (cars parked for an excessive duration) and at promoting soft modes in the city centre.

\paragraph{Users' satisfaction}
The Interview states that residents first questioned the need to deploy an SP system, but the system is now well established in the city. Regarding the smartphone application, the initial version was not satisfactory from a technical viewpoint and a new one is currently being integrated into the broader-scope city application.

\paragraph{Personal opinion on perspectives} 
Since its first deployment, the SP system has substantially evolved. Vendors, the policy officer in charge, as well as the dominant sensor technology have changed, but the project goes on, with a revised focus on cost-effective SP monitoring strategies (which enable to monitor a large number of spaces with only one sensor). The IT Manager sees no point in proposing a parking space reservation service in the future and he thinks this would be quite hard to implement on public on-street spaces.

\subsection{Grand Paris Seine Ouest, South West of Paris, France}
\subsubsection{City Context and Launch of the Smart Parking Project}
\emph{Grand Paris Seine Ouest} (GPSO) is an intermunicipal structure gathering 8 towns in the affluent South West periphery of Paris, namely, Meudon, Boulogne-Billancourt, Sèvres, Chaville, Issy-les-Moulineaux, Marnes-la-Coquette, Vanves, Ville-d'Avray, and 320,000 residents. Compared to other similar structures, GPSO has a wide range of competences, in particular in relation to all types of (paid) parking.

Overall, it manages 22,000 on-street spaces, as well as 5,000 spaces in off-street car parks. Non-residential on-street parking rates generally compare favourably with off-street ones; there are a few different parking zones, with hourly rates ranging from €1 to €2 and maximal durations above which the cost abruptly soars. Parking is free from 7pm onwards. For parking management, GPSO signed two public service delegation contracts for parking; the bid led by \emph{INDIGO} company was selected. While parking control and enforcement is still performed by agents in the field, GPSO launched an SP initiative in 2022-2023, with an operational start in 2024, to monitor the availability of on-street parking spaces, in addition to the already existing solution for off-street parking. This pilot study is conducted in two different towns : 567 on-street spaces were equipped in Issy-les-Moulineaux, along with 265 spaces in the centre of Sèvres, pending evaluation and possible extension. The parking information thus gathered feeds a dashboard used by technical services and city officials, and it is also displayed on parking guidance boards; no dedicated smartphone application has been developed.

\subsubsection{Insights from Mr. Hugo LAMBERT, Deputy Director for Mobility at GPSO}

\paragraph{Motivations}
 Mr. Hugo LAMBERT, Deputy Director for Mobility at GPSO, sees two main motivations to the implementation of on-street SP. Firstly, there is strong political will to make city centres easily accessible and to have a high turnover of parked cars, so that patrons can quickly find a space to park. This is partly inspired by the ``no parking, no business'' motto and the fear that patrons would favour department stores and their large parking lots otherwise. Besides, some towns in GPSO have a reputation for actively promoting innovation, notably smart-city technology. Secondly, he says that there is significant parking tension on the territory; in particular, on Wednesdays and Saturdays, outdoor market days, it is very difficult to park on the curb. Given that drivers prefer parking on street, informing them might spare them some cruising time and possibly redirect them to an off-street car park if they realise that on-street spaces are saturated.

\paragraph{Sensors}
Sensors are comprised of both ground sensors and overhead cameras mounted on streetlamps; they transmit parking information every five seconds. Camera-based systems are increasingly favoured, because they ``cover more parking spaces (if there is no visual obstruction), they are more reliable over time and, at the end of the day, cheaper''. As far as the Interviewee can tell, detection accuracy has not been a problem and the weather has not raised specific issues (but there have not been particularly severe weather conditions since their installation). Incidentally, he notes that around Christmas time, some anecdotal issues were encountered, because the cameras were slightly moved while Christmas street lighting was set up, which required new calibration. He also mentions that he has heard of other cities which have experienced parking detection with CCTV surveillance cameras; problems arose when their focus was changed for surveillance reasons, but this represents only a small fraction of time.  \\
By contrast, Mr LAMBERT deems that first-generation ground sensors (those released around 2010) yield less reliable data and that, in any event, their battery needs to be changed at some point and they may occasionally get damaged after being driven over so frequently''. There are still a number of ground sensors in the GPSO SP implementation, but their batteries are gradually changed for longer-life batteries. \\

\paragraph{Information display} Real-time parking information is conveyed to drivers  in the city centre via sign boards mimicking those for off-street parking; four levels of parking availability were defined based on arbitrary occupancy thresholds and for each parking area  the corresponding board shows the current level based on the data from the sensors. The Respondent expressed some doubt as to how easily they could be read in real time and mentioned that a revised design could better do the job.

Technical services also have access to parking information a centralised digital dashboard, updated every few minutes.
As parking enforcement still lies in the hands of agents on the ground, which echoes ``a strong political will'' and ``public disapproval of ANPR cars'', the SP dashboard plays a less central role than in other contexts (see, e.g., the case of Penang) and the server may therefore be used less intensively, but no connection issue has been reported since the start of the project.

On the other hand, GPSO has no plan to release a dedicated smartphone application at present. ``We do not want yet another application, everybody is tired of having 50 applications on their smartphones'', Mr. LAMBERT explains.

\paragraph{Impact on traffic}  To the best of the Respondent's knowledge, the traffic situation was not accurately diagnosed before the start of the pilot SP study. However, a quantitative analysis of the impact of the SP pilot was upcoming at the time of the interview, in which the parking situation (notably occupancy and turnover) in the first weeks after the start of the project (in lieu of ``just before'') would be compared with the situation at a later stage, with an established SP system. In parallel, questionnaires would help gauge the drivers' awareness of it and get their feedback.

\paragraph{Personal opinion and perspectives} 
The Respondent has underlined the importance of the political decisions which, on the one hand, have promoted innovative action on the path towards a smart city and, on the other hand, intend to preserve `manual' parking enforcement by agents on the ground, who are visible to citizens and can inform them. He admits that this choice is easier to justify in a territory like GPSO where illegal (i.e., not paid) parking is limited. \\
There is also a balance to strike between informing the drivers about the availability of parking spaces and urging them to drive to car parks, instead of cruising through the streets. Off-street parks are indeed generally much less occupied than on-street spaces and not more expensive, but drivers tend to prefer on-street spaces. From that perspective, GPSO's choice to display the saturation levels of different on-street parking areas, i.e., at an aggregate level, on variable-message signs makes sense, while going further and helping drivers identify vacant on-street spots in real-time may backfire and strengthen the drivers' tendency to look for on-street parking. More generally speaking, Mr. LAMBERT is convinced of the need to incentivise walking and the use of public transportation, but this should not be framed as a conflict between modes that takes place at the drivers' expense.

\subsection{Frauenfeld, Canton of Thurgau, Switzerland}
\subsubsection{City Context and Launch of the SP Project}

The town of Frauenfeld is the capital of the Canton of Thurgau in Switzerland and has a population of roughly 25,000.  Its transportation network consists of a railway station, two connections to the A7 motorway, and (for urban transport) 5 bus lines. Regarding parking, the town features 4 blue zones mainly dedicated to residential parking (with parking permits) and different areas for short-term (up to 30 minutes or 1 hour), medium-term (up to 2 or 4 hours), and longer-term parking. The parking rates are around CHF$\,1$ (i.e., €1.1) per hour and can be paid at the parking meter or on the \emph{ParkingPay} mobile application. In addition, there are 14 outdoor and indoor parking lots, among which 5 or 6 are owned by the town.

In November 2023, Frauenfeld launched a pilot SP project  to better understand and manage parking demand in its historic town center. The pilot covered about 500 parking spaces in mostly outdoor car parks. After its evaluation in 2024, the project was continued and  now relies on 24 overhead cameras (including CCTV cameras) mounted on facades, streetlamps, and even church towers.

\subsubsection{Insights from Peter KOCH, Director of the Communication and Economy Service) and Andrea FOSSATI, CEO of Parquery AG}

Our interview with Mr. Peter KOCH, Head of the Communication and Economy service at Frauenfeld, and Dr. Andrea FOSSATI, Co-Founder \& CEO of Parquery AG,  provided additional insight into this use case.

\paragraph{Motivations} 
Mr. KOCH indicated that ``the project goal was never to provide live steering of traffic, but rather to give the public reliable information and for the city to gain insights into actual parking behaviours''. Furthermore, the SP programme is not dedicated to parking control and no information is communicated to the police to enforce parking regulation. Consistently with this idea, no licence plate recognition is performed and the town only receives information about parking occupancy and duration, and entry and exit statistics. Finally, the municipal services (in particular the construction and traffic services) ``currently see no need for \emph{live} data''.

\paragraph{SP technology} The official argument for the technological choice is that cameras were used, and not ground sensors, to avoid high installation and maintenance costs associated with the latter, which have to be installed in each individual parking spot. To maximise the benefits drawn from each camera, the locations were chosen for their ability to capture occupancy data from multiple parking spaces. \\

Privacy concerns seem to have been carefully considered and privacy was built in the design of the system, to ease potential residents' concerns, thanks to low image resolution and high camera positioning. Once the locations of spots have been defined (thanks to \emph{Google Maps}), images are captured every 2 minutes (which results from a ``technical trade-off between power consumption'' and accuracy), processed in the cloud, and continuously deleted thereafter; the city only receives numerical parking data. Each camera is equipped with a battery, to overcome the periods of time when public electricity is cut. \\

Regarding the accuracy of the detection, the Interviewees answered that, once properly trained, the system achieves ``above 99\% accuracy'' in general, but is more challenged if the camera captures a distant view of the scene, in which case specific re-training may be required and the accuracy decreases to 95\%. Thus, the position of the camera is the feature that impacts accuracy most. The weather is not critical, except when there is thick fog, and the system is operational at night, thanks to street lighting.

\paragraph{Data availability}
The parking data collected every two minutes are made available to the public via their integration in the 
\emph{Regio} Frauenfeld mobile application 
\footnote{This application is written in German and does not provide translation to any other language, which may limit its use by visitors.} and Canton Thurgau open data platform \footnote{\url{https://data.tg.ch/explore/dataset/frauenfeld-2}}. Statistics on occupancy rates, parking durations, entries, and exits, are thus easily accessible. Integration with the most common navigation applications is an ongoing effort.

\paragraph{Financial aspects}  Financially, the project was a significant but manageable investment. The cost of approximately CHF$\,$100,000  was equally shared between the City of Frauenfeld and the Canton of Zurich and annual operating expenses may amount to about CHF$\,$30,000. (In comparison, boards to display the information come at a cost of about CHF$\,$1,000,000.)

\paragraph{Actual benefits and impact on traffic} While the impact on traffic or search times was not checked and no survey was conducted to get public feedback, the Interviewees' impression is that SP has had a positive impact. Above all, they noted that ``political leaders gave positive feedback, recognizing the system’s contribution to making city life easier and communication more transparent''. They did not hear of any dissent based on the argument that the system might promote car use by improving convenience, rather than encourage a modal shift toward public transport.

From the perspective of drivers, no frustrations were reported. Citizens understood that the system was designed as a "parking information system" rather than a "guidance system", and therefore did not expect perfectly live, always-accurate data. While external factors such as camera angle, daytime, and weather occasionally influenced detection accuracy, overall performance was considered robust and reliable.

\vspace{0.5cm}

Finally, for the last two case studies, we turn to cities that have pondered on-street parking management strategies, but eventually chose not to implement on-street SP systems.

\subsection{Zurich, Switzerland}

\subsubsection{City Context and Parking Policy}

Zurich, Switzerland’s economic hub and largest city with a municipal population of 440,000, is known for its efficient public transport and progressive urban policies. Daily mobility flows are intensive, insofar as there are around 400,000 commuters every day (including intra-city commuters), more than 60\% of whom come from outside the city boundaries  \cite{zurich2013bewegung,zurich2022bewegung}. Over the decades,  mobility flows have drastically changed: while about half of the commuters used their private cars in the 1980s, the private-car share sunk to 26\% in the last decade \cite{zurich2013bewegung,zurich2022bewegung}, and even less for intra-city commuters. Meanwhile, 62\% of commuters relied on public transportation.

 The restrictive parking policy that has been implemented for several decades has played a role in this shift \cite{kodransky2010europe}, with the introduction of parking \emph{maxima} and the removal of on-street parking from residential areas. As of 2019, the city had 42,200 public street parking spaces (a lower figure than in 2010 despite the population growth), the majority of which are in ``blue'' zones mostly for residential parking. An additional 25,700 spaces are  accessible to the public in parking lots and car parks \cite{zurich2025stadtverkehr2019}. The trend was pursued between 2022 and 2025, with the dismantling of on-street parking spaces and improvements targetting bicycles and pedestrians, as well as enhanced green spaces. It also directly follows from these policies that the city is widely reported to face high parking tension. 

The City implemented a real-time parking guidance system to 36 of its largest and busiest parking facilities, with over 200 on-street signs guiding drivers to nearby facilities. For on-street parking, there is no such smart-parking equivalent. Peak on-street parking demand is between 11:00–13:00 on weekdays and in a broader time range on Saturdays in the city centre, and on-street  parking rates are set on a static schedule:  CHF 4 (about €4.4) per hour in the city centre, with maximal parking duration between 30 minutes and 2 hours, typically. These rates have to be compared with the corresponding off-street parking rates, which typically amount to CHF 4 for the first hour and CHF 5 for every following hour (even though there are longer-term options); parking at night and on Sundays is cheaper.

\subsubsection{Insights from Mr. Georg SPYCHER, General Manager, Parking Zurich AG}

\paragraph{Standpoint on parking-related measures}
Mr. SPYCHER (the General Manager of \textit{Parking Zurich AG}) admits from the outset that drivers in Zurich often express frustration, but the city has intentionally avoided enhancing private car convenience: its policies are designed to discourage car ownership and instead promote public transport and shared mobility. ``Looking forward, Zurich plans to further adjust parking fees and reduce maximum parking durations to continue promoting sustainable urban mobility'', he said.

\paragraph{Vision on urban mobility}
In Mr. SPYCHER's vision,
``parking must evolve to accommodate autonomous and self-driving cars, while also ensuring accessibility for people with disabilities. Car sharing, in particular, is a positive step as it reduces congestion and the overall demand for parking''.  A prime example of Zurich’s approach is the Parkhaus Zürichhorn, which balances 220 spaces among 240 tenants with 30–70 spaces dynamically available public spaces, optimizing usage across different groups rather than simply expanding capacity.

Parking in Zurich is therefore not just about cars and spaces; it reflects a deliberate strategy to shape urban mobility. Payment remains simple and structured, while public transport is prioritized as the backbone of movement across the city. The message is clear: Zurich’s future lies in shared, sustainable, and technologically adaptive mobility rather than in accommodating more private vehicles.

\subsection{Perth, Australia}
\subsubsection{City Context}

The metropolitan area of Perth,  the capital of Western Australia, is home to  over 2 million inhabitants. A vibrant commercial and cultural hub, the city of Perth has  municipal population of 110,000 residents, but an estimated 200,000 employees, students, and visitors commute into the city centre every day \cite{perthCommutingFlows}. To support this pulse of activity, the city manages a significant on-street parking inventory, with over 6,000 pay-and-display bays across the central district, in addition to multiple car parks managed by commercial entities or by Perth. However, to cope with the commuting flows, since 2010 the transport planning authorities have officially endeavoured to restrict parking in the central area and promote public transportation and/or park-and-ride for commuters
\cite{perthCommutingFlows}. In parallel, parking meters have been modernised and ticketless pay-by-phone solutions have been deployed. Parking control by automatic number-plate recognition (ANPR) is also enforced at car parks.

Pricewise, the first 15 minutes of parking are free. Then, the on-street parking rate depends on the zone, with a rate of AU\$7.00 (about US\$4.60) per hour in the central business district (CBD) and AU\$3.00 in Crawley/Nedlands. Maximum time limits (typically, 1 hour or 2 hours) are displayed on dedicated kerbside signage, but there also exists an online interactive map\footnote{ \url{https://perth.maps.arcgis.com/apps/webappviewer/index.html?id=29d9a05a4f394f02a89084997e11566e}} to find parking-related information. Recently, hourly and daily rates rose modestly (by AU\$0.20 to AU\$0.60 per hour) and  a AU\$5.00 flat night fee was reinstated in place of free night-time parking.

\subsection{Insights from Justin McKIRDY, Executive Director of Urban Mobility with DTMI, Western Australia}

Right from the start, Justin McKIRDY, Executive Director of Urban Mobility with DTMI in Western Australia, highlighted
that different levels of responsibility ought to be distinguished on the topic of parking, even though they are entangled. First, the role of the metropolitan transport authority (to which he belongs) should not be mistaken for that of the city, which operates on-street parking. Second, parking policy --  which results from an agreement between the city and the State in the frame of the now 24-year-old Parking Management Act -- is partly disconnected from regulation (to control parking locally). Third, parking measures differentiate between different parking types (residential/tenant parking, etc.).

\paragraph{Differentiated parking tools and goals} 
Besides the contactless payment option (and the camera-based ANPR in place at the gates of off-street car parks), no other SP strategy has been implemented for on-street parking. Instead, according to the Respondent, the major goal is to keep excess traffic out of the city centre as much as possible, notably by capping the parking supply; parking management is particularly critical in the ``corridors' leading into and out of the CBD core area. Parking in this area should be mostly on a short-term basis (as prompted by the free first-15-minute measure) between peak hours, or even on a very short time basis in some specific bays. Incidentally, the Respondent does not have much information 
about the impact of the free 15-minute policy, given that the related policy is managed by the City of Perth. As for commuters, who require longer-term parking, Mr McKIRDY stated that it is desirable that they be intercepted at the periphery and encouraged to walk or use public transport.

\paragraph{Parking occupancy and traffic}
Given the necessarily limited parking supply, on-street parking in the CBD is well used, but not to the point that drivers ``queue in front of these [spaces]''. Parking occupancy values are in the hands of the City of Perth\footnote{Brief releases indicate that surveys are commissioned annually to assess parking occupancy; their latest edition reports occupancies typically between 40\% and 80\%, for car parks and street parking indiscriminately, with only a couple of oversaturated areas.}, and not the Department of Transport,  but the Respondent is not convinced of the accuracy of these data. Since underused parking bays may be an argument for reducing the City levy, there might also be risk of a bias towards occupancy underestimation.
Still, even though traffic is congested at peak hours (a couple of hours in the afternoon, in particular), the parking restrictions have borne some fruit, preventing a surge in traffic in the CBD.

\paragraph{Main levers of action} Occupancy signage for car parks exists in the CBD, but Mr. McKIRDY is not sure that this plays a major role. By contrast, there is anecdotal evidence that drivers are sensitive to the cost of parking, insofar as the cheaper or free parking options are filled first (e.g., the free East block before the paying West block).

\paragraph{Upcoming challenges} Rather than technological innovations facilitating parking search, the Interviewee reckons the the major upcoming challenges are the acceptance and adaptation to the revised parking legislation and the public wish to contain traffic in the CBD in spite of paramount demographic changes in the years to come. Indeed, the CBD is ``transitioning from a business-oriented area to a mixed-use area'', with projections of a massive surge in the number of residents, arriving at a density that is new for Perth. In parallel, business is also expected to intensify in the CBD. These two evolutions will tend to generate significant car traffic and require more parking spaces, even as the transport authorities strive to keep traffic out of the CBD.
\\
Looking forward, the city is steering mobility with a broader vision. ``Transitioning and growing populations require more dynamic and sophisticated solutions for sustainable public transport, the city is working in that direction''.

\section{Discussion and Conclusion}\label{discussion}

In summary, our survey of actual city implementations of on-street SP shed light on a diversity of contexts, technological choices, deployment scales, and project viabilities. Ground sensors were used in widely mediatised pilots in the early 2010s and are still used in more recent, larger-scale implementations, but since 2018$\sim$2020 some cities have gradually shifted to static cameras with computer-vision capabilities, which enable them to monitor more parking spaces per device and may be more durable. In parallel, a growing number of cities are resorting to mobile cameras aboard ANPR cars for parking control, which provide extensive spatial coverage, at the expense of a low temporal resolution; but only a fraction of them have overtly decided to turn the data thus collected into a \emph{bona fide} SPS with real-time parking availability information communicated to technical services and/or drivers. In terms of market stage, recent years have witnessed an apparent consolidation of the SP industrial ecosystem, with several mergers between companies.

We complemented this objective survey with various use cases around the world, asking stakeholders for an open feedback; this sometimes brought some nuances to the publicised ``success stories'' that can be read online.  What emerges from these discussions is that contexts and situations dramatically differ, not only with respect to city sizes and chosen SP implementations, but also regarding the main goals of the SP projects. In this section, we provide actionable insight for researchers and practitioners by discussing the extent to which these diverse goals have been achieved in practice and by highlighting some points that require particular attention for an SP project, in light of our case studies.


\subsection{Reduced Congestion and Parking Search Times in City Centres}

Cruising for parking is often mentioned as a substantial contributor to congestion in cities, with popular figures for the share of cruising traffic over total traffic in the 20–30\% range  \cite{shoup2011parking,shoup2006cruising}, even though recent trajectory-based evidence suggests to take reports of very long city-wide \emph{averages} of search times (say, longer than 5 minutes) with a grain of salt \cite{saki2024cruising}. Does SP \emph{actually} manage to expedite parking search?

The best documented evidence in support of this assertion comes from the pioneering \emph{SFpark} project in San Francisco (2011-2014). Thanks to a data-based dynamic adjustment of parking rates, improvements in parking availability were achieved, with measured occupancies in pilot areas that \emph{more often} met the 60\% to 80\% target range (+31\% of the time, and up to +100\% of the time on block faces with high payment compliance); reported search times were also reduced (by 5 minutes in pilot areas, vs 1 minute in control areas, where the parking tension was already less acute beforehand) \cite{SFPark2014evaluation}. That being said, it is also critical to remember that \emph{SFpark} was a pilot study that covered only a part of the on-street spaces in the city centre and that was terminated 3 years after its start (hardware issues emerged about one year into the project).

The case of Penang gives an example of a larger-scale, city-wide, and durable deployment, which did not use the data for parking rate adjustments. While there are claims of substantial reductions in  CO\textsubscript{2} emissions and parking search times (by 30\% and 50\%, respectively) thanks to SP in Penang, the evidence supporting these claims is scant. For other implementations, on much smaller scale, stakeholders did not observe visible changes in traffic;  note that awareness of the SP solution is critical in this case, particularly if the transport authority hopes to influence drivers' behaviour in an educational way, by \emph{informing} them about the scarcity of available on-street parking spaces, e.g. along the Passe-A-Grille beach in St Pete Beach.

More generally, rigorous evaluations, measuring before–after changes in cruising, search times, or emissions with standardised methods, remain scarce. In the vast majority of our case studies, a \emph{quantitative} diagnosis of the traffic and parking search conditions before (and possible after) SP implementation is missing. In short, SP offers several levers for 
\begin{itemize}
    \item reducing search times, by displaying real-time parking information on sign boards or on a mobile application, guiding drivers to spaces that are likely to  be vacant, or enabling them to reserve a space in advance (an option hardly considered for on-street parking), 
    \item soothing the cruising traffic, using either a passive strategy, i.e., just informing drivers about the current parking tension in busy areas and hoping they will choose to park elsewhere, like in St Pete Beach or GPSO, or a more interventional one, such as adjusting parking rates, like in San Francisco. 
\end{itemize}
But, in practice there are but few actual measurements of reductions in congestion and parking search times to this day.

\subsection{Increased Parking Revenue for Cities}

Revenue and cost-effectiveness must be judged on lifecycle economics, balancing SP installation, operation, and maintenance costs against possibly higher fee collection and lower enforcement costs.
Interestingly, in our case studies, city councils did not have to substantially invest for SP deployment in general, as the system deployment and sensor installation costs were often funded by a national or federal grant, or bundled into a monthly subscription fee, or taken up by the vendor/service provider in exchange for shared parking revenue (e.g., in Penang).  Capital risk is thus shifted off the city balance sheet, but the practicalities of revenue sharing are of paramount importance, as demonstrated by the Penang and Douala examples.

Parking revenue was reported to increase slightly in the \emph{SFpark} pilot, mostly due to the smart payment option and longer allowed parking durations, but by far not sufficiently to cover sensor replacement and maintenance without new budgetary arrangements. Typically, one expects SP to increase parking revenue by enhancing compliance, when it is low. In this vein, 
the small-scale Douala pilot exhibited a dramatic increase in parking revenue. On a larger scale,  Penang fully relies on SP to collect parking revenue, which facilitates payment and control, but bears on the city's finances on the few days when the service is down.

In situations where no substantial increase in revenue is expected, material costs come into play in a more critical way, and many of our Interviewees said camera-based systems were favoured over ground sensors because of their higher cost effectiveness, as they cover multiple spaces and may be more durable.

\subsection{Data-based Pricing Strategies To Tweak Occupancy and Turnover Rate}

Changing the parking rate is the most straightforward lever to control curb occupancy and turnover. The pricing strategy can be applied at a very aggregate level, e.g., by setting high parking rates  to disincentivize private car use. For instance, this was part of Zurich's long-term policy, along with the limitation of the on-street supply and investments in public transport; it effected a profound modal shift, with an estimated 17\% decline in automobile ownership from 2000–2017, illustrating the potential of pricing combined with broader mobility strategy. At an intermediate scale, when rates vary with the area, for instance when the city is divided into a few zones associated with different parking conditions and colours (e.g., red/orange/yellow/green in Montpellier), turnover rate and occupancy can be modulated in space \cite{dutta2021searching} (e.g., to promote very high turnover on red parking spaces in Montpellier). These pricing strategies, operating at an aggregate level, are not data-intensive, in that they require only time and space averaged occupancy and turnover data.

On the other hand, dynamic pricing at the scale of the block-face requires finer-grained measurements, hardly achievable without SP. SFpark operationalised this idea by adjusting rates every eight weeks to reach target occupancies (e.g., 60–80\% target ranges on 2 opposite streets Lombard \& Chestnut) and observed improved availability in pilot blocks relative to controls. The effectiveness of this strategy hinges on visible, predictable rules, and considerable citizens' awareness:  alongside high-frequency occupancy monitoring, clear political communication is crucial. Also, block-level variations in parking rates may run counter to the local habits, for instance in Europe, where such spatial heterogeneity is unusual. Besides, price adjustments must be calibrated based on the local price elasticity, which may vary considerably at the block level \cite{shoup2011parking}.

\subsection{Drivers' and Citizens' Satisfaction}

Regarding the citizens' satisfaction, a first concern resides in the installation of the sensors. Here, too, the local context is key. In some contexts, SP sensors have not raised particular concerns: no such concern was voiced by our Interviewees in San Francisco, Douala, Penang. In others contexts, considerable efforts have been made to avoid giving citizens the impression of encroachments on their privacy, notably in Frauenfeld and St Pete Beach, where this issue guided the choice of sensor technology. In some sites (including St Pete Beach), aesthetic concerns associated with the visibility of the sensors have also played an important role.

Turning to drivers' satisfaction, obviously, price is a major concern. If the SP deployment coincides with a general hike in parking rates, dissatisfaction can be expected (and this was carefully avoided in the San Francisco pilot, by refraining from systematically raising rates).
Then, real-time SP data are only beneficial to drivers that are aware of, and use, these data. In this respect, limited awareness of SP information \cite{peng2017impacts} is often a critical problem, especially in small-scale implementations. Incidentally, while reviewing implementations globally, we phoned unknown shop owners and service managers in areas where SP had been deployed, notably in Pozuelo de Alarcon, near Madrid, Spain, and found that they were not aware of any SP deployment. In a similar vein, in the situations where a mobile application was developed to display real-time parking availability, the application may be hard to identify. In many cases, particularly small-scale ones, the original application was  discontinued, either because the vendor went bankrupt (see the case of \emph{ParkingMaps} in Soissons) or the application changed names, possibly after integration into a broader application (see the integration of Mosman (Australia)'s SP application into New South Wales' \emph{Park'nPay}). Alternatively, there may be multiple competing applications, among which it is hard to find which ones feed on live SP data (e.g., on the data communicated by the sensors in Los Angeles, USA).

By contrast, the case study of Penang gives an example where one specific mobile application is widely downloaded (over 1.5 million times, according to official sources) and used -- notably, but not exclusively, to pay for parking. Still, even in this context of broad SP awareness, some dissatisfaction may be voiced, for instance users complaining about device compatibility and interface problems, along with server failures now and then. Public acceptance of SP assistance may be easier to garner if the system is presented as an \emph{information} service, as for Frauenfeld’s camera system, rather than an indispensable, be-all-and-end-all parking utility.

Despite their investment in SP, some cities (see the case study on GPSO) decided to turn away from the development of a mobile phone application and to favour the physical installation of dynamic sign boards that display live information about parking availability at an aggregate scale (see GPSO, Soissons). Stakeholders in these cases argue that there is widespread fatigue with the multiplicity of smartphone applications. On the other hand, stakeholders in other contexts (e.g., Perth) have the feeling that such signage has little influence on drivers' routing choices.

Last but not least, trust in the SP-related information is naturally vital. The St Pete Beach case study hinted at the skepticism of some users about the real-time parking availability data. Press clippings point to even larger skepticism during the short-lived SP project in Nice, France. With regard to reliability, inaccuracies due to the hardware may rightly be in the firing line, as field experience has evinced degrading performance of ground sensors, in particular, over time scales of one or two years: sensor lifetimes are shortened by battery depletion, street works and harsh environments; for example, \emph{SFPark} experienced progressive sensor failures that curtailed long-term benefits after the pilot phase.  But it might also be that, while someone was driving to a spot marked as vacant on the application, another driver pulled in in the meantime, which is equally frustrating for the driver.

\subsection{Points of attention and vulnerabilities}
Beyond the foregoing strengths and limitations with respect to the attainment of SP goals, we would like to conclude by underlining some points of attention that emerged from the review and case studies.
Clearly, SP systems are neither a silver bullet nor a trivial add-on, but complex socio-technical infrastructures that succeed only when reliability, pricing, and governance are aligned. As such, smaller-scale implementations and pilots are particularly vulnerable to changing political will, discontinued funding, or defaulting start-ups.
The regional context also matters: if an SP project blooms on propitious ground, i.e., if SP seems to be on the rise regionally (e.g., in New South Wales, Australia, or in the United Arab Emirates), regional integration of the SP application can be contemplated. Alternatively, \emph{functional} integration into a broader-scope application (not just for parking information and smart payment, but also for other city services) is a means to gain visibility and overcome the application saturation problem.

In any event, for practitioners, the emphasis should be on building systems that can persist in time. Besides the sensor technology, the financial model also matters in this regard: sustainable maintenance budgets or well-structured public–private partnerships are prerequisites to keeping systems trustworthy over time. Above all, occupancy data must translate into actionable policy, whether through dynamic pricing, enforcement, or communication. Otherwise the technology risks delivering visibility without impact.

\section*{Acknowledgments}
We acknowledge financial support by \emph{Transports de l'agglomération de Montpellier} and from CNRS through the MITI interdisciplinary programs.

\section*{Declaration of competing interests}
Funding was received from \emph{Transports de l'agglomération de Montpellier} for the analysis of their smart parking data and assistance in the development of their guidance algorithm. The associated contract explicitly mentions that the parties can freely communicate and publish their findings.

\section*{Declaration of generative AI and AI-assisted technologies in the manuscript preparation process}
Conversational AI (ChatGPT) was used in the first stage to identify resources related to smart parking implementations and to flesh out related paragraphs. Then the references and the numbers were all manually checked by the authors and the final version of the manuscript was written by the authors. The interviews were conducted and summarised without resorting to AI.
\newpage


\end{document}